\documentclass[fleqn,usenatbib,useAMS]{mnras}

\usepackage{graphicx}	
\usepackage{amsmath}	
\usepackage{amssymb}	
\usepackage{multicol}        
\usepackage{bm}		
\usepackage{pdflscape}	

\usepackage[T1]{fontenc}
\usepackage{ae,aecompl}

\usepackage{newtxtext,newtxmath}

\newcommand{\bz}{\ensuremath{\langle B_{\rm z} \rangle}}
\newcommand{\bs}{\ensuremath{\langle \vert B \vert \rangle}}

\title[A test of the unipolar inductor for GD\,356]{A test of the planet-star unipolar inductor for magnetic white dwarfs}

\author[N. Walters et al.]{N. Walters,$^1$\thanks{E-mail: nikolay.walters.15@ucl.ac.uk}
J. Farihi,$^{1}$
T. R. Marsh,$^{2}$ 
S. Bagnulo,$^{3}$
J. D. Landstreet,$^{3,4}$
J. J. Hermes,$^{5}$ 
\newauthor{N. Achilleos,$^{1,6}$} 
A. Wallach,$^{5}$
M. Hart,$^{5}$ 
C. J. Manser$^{2}$ 
\medskip
\\
$^1$Department of Physics and Astronomy, University College London, London WC1E 6BT, UK\\
$^2$Department of Physics, University of Warwick, Coventry CV4 7AL, UK\\
$^3$Armagh Observatory and Planetarium, College Hill, Armagh BT61 9DG, UK\\
$^4$Department of Physics and Astronomy, University of Western Ontario, London, Ontario N6A 3K7, Canada\\
$^5$Department of Astronomy, Boston University, 725 Commonwealth Ave., Boston, MA 02215, USA\\
$^6$The Centre for Planetary Sciences at UCL/Birkbeck, London WC1E 6BT, UK
}

\date{Accepted XXX. Received YYY; in original form ZZZ}

\begin{document}


\maketitle

\begin{abstract}

Despite thousands of spectroscopic detections, only four isolated white dwarfs exhibit Balmer emission lines. The temperature inversion mechanism is a puzzle over 30 years old that has defied conventional explanations. One hypothesis is a unipolar inductor that achieves surface heating via ohmic dissipation of a current loop between a conducting planet and a magnetic white dwarf. To investigate this model, new time-resolved spectroscopy, spectropolarimetry, and photometry of the prototype GD\,356 are studied. The emission features vary in strength on the rotational period, but in anti-phase with the light curve, consistent with a cool surface spot beneath an optically thin chromosphere. Possible changes in the line profiles are observed at the same photometric phase, potentially suggesting modest evolution of the emission region, while the magnetic field varies by 10 per cent over a full rotation. These comprehensive data reveal neither changes to the photometric period, nor additional signals such as might be expected from an orbiting body. A closer examination of the unipolar inductor model finds points of potential failure:  the observed rapid stellar rotation will inhibit current carriers due to the centrifugal force, there may be no supply of magnetospheric ions, and no anti-phase flux changes are expected from ohmic surface heating. Together with the highly similar properties of the four cool, emission-line white dwarfs, these facts indicate that the chromospheric emission is intrinsic.  A tantalizing possibility is that intrinsic chromospheres may manifest in (magnetic) white dwarfs, and in distinct parts of the HR diagram based on structure and composition.

\end{abstract}

\begin{keywords}
	stars: individual: GD\,356---
	stars: magnetic field---
	stars: white dwarfs---
	starspots---
	planetary systems---
	planet-star interactions
\end{keywords}

\section{Introduction}

In recent years, there has been a rapid rate of white dwarf discovery, ensuing from the advent of modern large-scale surveys. Currently, over 800 white dwarfs have been identified as magnetic (e.g.\ \citealt{Kepler2015, Kepler2016}), with field strengths ranging from as high as 1000\,MG and down to a few kG, below which successful detection of magnetism via spectropolarimetry is challenging \citep{Landstreet2012}. From studies of the 20\,pc sample \citep{Holberg2016,Hollands2018}, it is estimated that at least 12\,per cent of white dwarfs are magnetic. This estimate is comparable to the conclusion of an earlier study based on spectropolarimetry of hydrogen-rich stars \citep{Kawka2007}, but recent work taking advantage of metal-lined white dwarfs has increased this estimate to $20 \pm 5$\,per cent \citep{Landstreet2019,Bagnulo2019,Bagnulo2020}. Three-quarters of the known magnetic white dwarfs are thought to be isolated, or are members of non-interacting binary systems \citep{Ferrario2019}. Research on magnetic white dwarfs that exhibit atypical properties remains crucial for the insight into the magnetic fields of this population. The origin and evolution of magnetism in white dwarfs is an outstanding astrophysical problem, and thus detailed studies of individual systems may shed light on their exact nature.

Two theories have been developed to account for the presence of magnetic fields in isolated degenerate stars: a fossil magnetic field \citep{Tout2004,Wickramasinghe2005} and convective dynamo field \citep{Tout2008,Potter2010}. An interplay of the two could also be a possible explanation \citep{Featherstone2009} based on the complex field structures inferred from time-resolved spectropolarimetric observations \citep{Kawka2020}. The fossil-field origin was originally proposed over half a century ago \citep{Landstreet1967}, and first tested via the distribution of magnetic field strengths in the population of white dwarfs by \citep{Angel1981}. A link was found between the magnetic fluxes of main-sequence Ap and Bp stars and high field magnetic white dwarfs, once field amplification due to evolutionary effects is taken into account. However, it has been shown more recently that the observed magnetic Ap-Bp star space densities are not sufficiently high to account for the fraction of strongly magnetic white dwarfs \citep{Kawka2004}. Additionally, a fossil field would have to survive the turbulence associated with the RGB and AGB phases of stellar evolution, although field lines may be trapped and retained in non-convective core regions. 

Despite these inconsistencies, a recent study found three magnetic white dwarfs identified as members of open clusters, and firmly linked to single, intermediate-mass star evolution \citep{Caiazzo2020}. This result supports the idea of fossil field origin in at least some cases. On the other hand, a strong magnetic field could be brought into existence during the common envelope stage of a low-mass companion and the giant progenitor of a white dwarf \citep{Nordhaus2011,Wickramasinghe2013}. In this case, smaller seed fields could be significantly amplified in the common envelope. Indeed, the dearth of magnetic white dwarfs with detached companions argues indirectly but strongly for this binary evolution channel for magnetic field production \citep{Liebert2005}.

The presence of a strong magnetic field can have a direct effect on the appearance of the stellar photosphere due to surface inhomogeneities, frequently manifested as photometrically inferred spots \citep{Brinkworth2005, Holberg2011,Hermes2017a,Reding2018}, as well as effects on accretion in binary systems \citep{Ablimit2019}, and convective mixing \citep{Tremblay2015}. Field-inhibited atmospheric convection can trigger the formation of dark spots \citep{Valyavin2014}, while a lower mixing rate between the outer layers and a convective envelope may result in a non-uniform chemical surface composition \citep{Ferrario2019}. A combination of stellar rotation and variation in surface composition can produce continuum flux and polarimetric modulation via magnetic dichroism \citep{Achilleos1992}, thus resulting in photometric variability. Moreover, spectroscopic changes in the Zeeman-split components may appear due to variations in the surface field strength through a rotation cycle \citep{Kilic2019}.

GD\,356 (=WD\,1639+537) is an isolated and cool magnetic white dwarf \citep{Greenstein1985} with fundamental parameters summarised in Table \ref{tab:Tab1}. Its highly unusual optical spectrum lacks absorption lines, but exhibits broad H\,$\upalpha$ and H\,$\upbeta$ emission features in Zeeman-split triplets, translating into a rare -- and until only recently unique -- DAHe spectral type. However, the spectral energy distribution is best fitted by a helium atmosphere model, and interpreted as such \citep{Bergeron2001}. Analysis of the emission line profiles of GD\,356 suggests a magnetic field described by  a centred magnetic dipole with a polar field of $13 \pm 2$\,MG \citep{Ferrario1997a}. This archetypal white dwarf exhibits brightness variations with amplitude $\pm0.2$\,per cent ($V$-band), and period $1.9272\pm0.00002$\,h that is interpreted as the white dwarf spin, where light curve modelling yields a polar (or inclined equatorial) starspot which covers 10\,per cent of the surface \citep{Brinkworth2004}. A similarly-sized spot is inferred as the source of emission lines in a temperature-inverted region based on spectropolarimetry and detailed modelling \citep{Ferrario1997a}. 

The exact physical nature of such a spot and its origin remain unclear, despite extensive and multi-wavelength efforts to detect a companion, a corona or accretion \citep{Ferrario1997a,Weisskopf2007,Wickramasinghe2010}. Interestingly, at least two, and possibly three  further examples of DAHe stars have been reported within the past year \citep{Reding2020,Gansicke2020,Tremblay2020}; all are isolated (i.e.\ not cataclysmic variables, no circumstellar disks), relatively fast rotators, and have cool atmospheres where $T_{\rm eff} < 10\,000$\,K. Thus GD\,356 is now the prototype of a small class of DAHe stars, but the new discoveries only deepen the mystery. The only published and still viable theory for the nature of GD\,356 is the unipolar inductor. In this model a conducting planet, orbiting within the magnetosphere, sets up a current loop that dissipates in the stellar atmosphere via Ohmic dissipation and thus heating the outer layers into emission \citep{Li1998}.

Motivated by the unipolar inductor model, this study aims to detect additional periodic signals from the influence of such a closely orbiting planet. This paper presents rotation phase-resolved spectroscopy and spectropolarimetry of GD\,356, which is then compared with the latest NASA \textit{Transiting Exoplanet Survey Satellite} \citep[\textit{TESS},][] {Ricker2015} data release, and ground-based, time-series imaging to investigate the spectroscopic and photometric variability. The observations are described in Section \ref{sec:obs}, and in Section \ref{sec:res} the analytic approaches and results are outlined. The implications of the results are then discussed in Section \ref{sec:dis}, followed by conclusions in Section \ref{sec:conc}.

\section{Observations and Data}\label{sec:obs}

A combination of optical photometry, spectroscopy and spectropolarimetry was obtained. See Table \ref{tab:Tab2} for details of these datasets which are detailed in the following sub-sections.

\subsection{Time-Series Spectroscopy}

\begin{figure*}
\includegraphics[width=\linewidth]{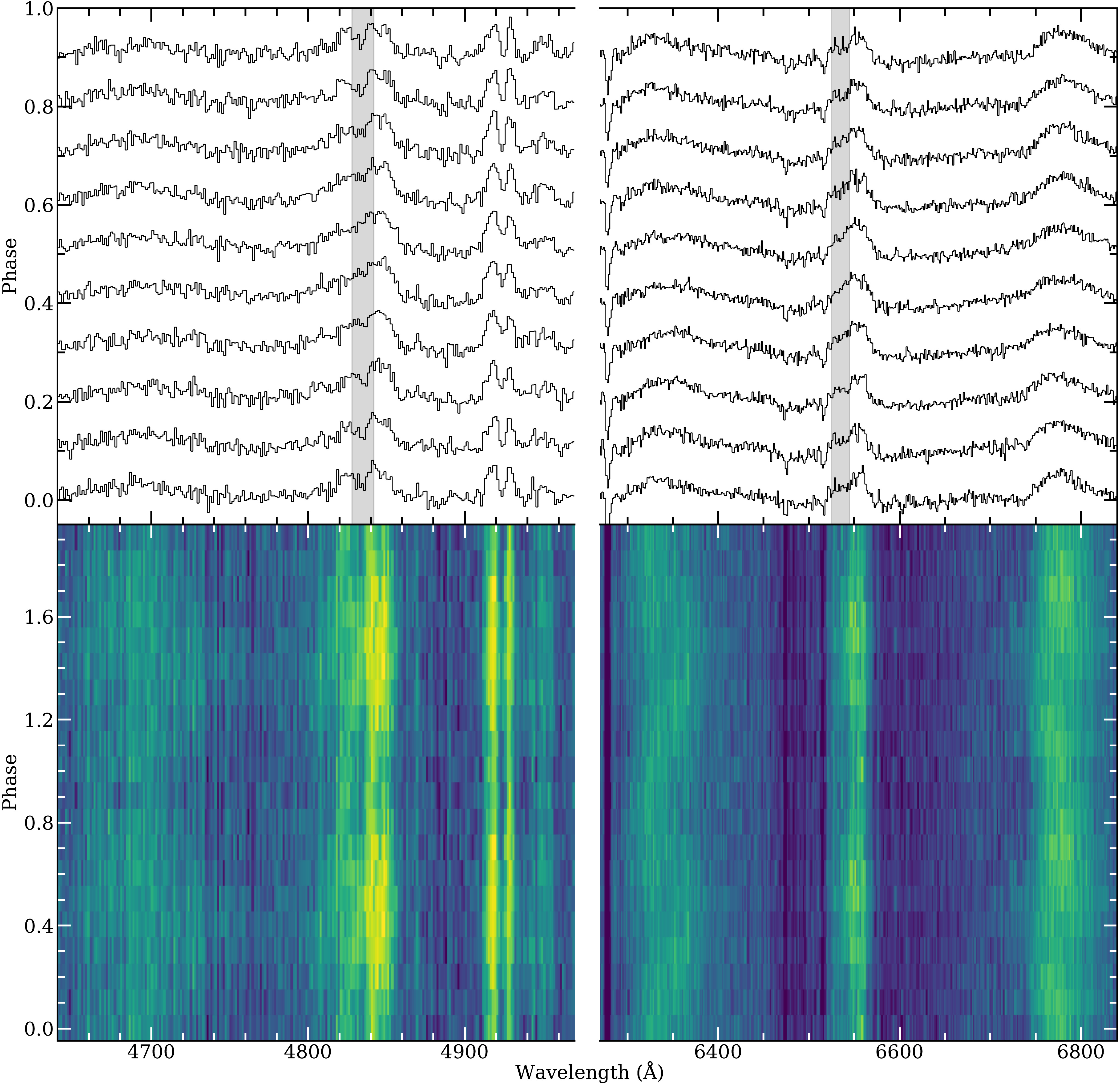}
\vskip 1mm
\caption{\textit{Upper panel}: Phase-folded H\,$\upbeta$ and H\,$\upalpha$ normalised ISIS spectra. The spectra were folded on the established period where phase $=0$ corresponds to photometric maximum at $\rm{BJD}_{\rm{TDB}} \approx 2458296.0255$. Shaded regions highlight (potential) periodic dips in both central $\uppi$ components, where this variation is clear in the H\,$\upbeta$ component, and less certain in H\,$\upalpha$. The absorption feature near 6375\,\AA \ is telluric O$_2$. The red arm data were re-binned to 1.5\,\AA, and blue arm data to 1.3\,\AA \ for better visualisation. \textit{Lower panel}:  The data from the upper panel plotted as a trailed spectrogram. The maximum relative flux is indicated in yellow and the minimum in dark blue. Periodic variations in both the relative emission intensity and central wavelength are visible in multiple features, with the latter most clearly evident in the $\upsigma_{\rm b}$ component of H\,$\upalpha$.}
\label{fig:Fig1}
\end{figure*}

Spectroscopic observations were carried out on the 4.2-m William Herschel Telescope (WHT) at Roque de los Muchachos, La Palma. Time-series data were obtained on 2018 July 29 and 2019 May 14 in service mode, using the Intermediate-dispersion Spectrograph and Imaging System (ISIS), a double-arm, low- to medium-dispersion, long-slit spectrograph. The instrument was used in the default service mode setup, including the EEV12 $2048\times4096$\,pixel$^2$ detector in the blue arm, and Red+ $2048\times4096$\,pixel$^2$ array in the red arm. On-chip binning of 3 pixels in the spatial direction was employed to increase signal-to-noise (S/N) in the case of potentially poor or variable seeing. A slit width of 1.2\,arcsec was used together with the R600B and R600R gratings centred at 4500\,\AA \ and 6500\,\AA \ respectively. This resulted in an unvignetted wavelength range of approximately $3920-5150$\,\AA \ in the blue arm with resolving power $R\approx1800$, and $5900-7100$\,\AA \ in the red arm with $R\approx2700$. Exposures were taken in both arms simultaneously, with individual integration times of 120\,s each. In this manner, both observing runs covered around 4.5 contiguous hours on-source, and yielded, on average, 116 frames of data in each arm. Such a setup ensured full spectroscopic coverage of H\,$\upalpha$ and H\,$\upbeta$ emission lines (Fig.~\ref{fig:Fig1}).

The target frames were obtained under bright conditions with lunar illumination of 96 and 81\,per cent during the 2018 July and 2019 May runs, respectively. Weather conditions during the 2018 July run were dusty, but cloudless, and seeing ranged from 0.5 to 0.8\,arcsec. The seeing in the 2019 May observational period varied from 0.8 to 2.0\,arcsec, with an average of approximately 1.0\,arcsec and no clouds. For both runs, the airmass ranged between 1.1 to 1.6. Multiple flat, bias, and arc lamp frames were taken on each night for calibration purposes, but only a single spectroscopic standard star was observed for the program (during the 2019 May run). These service-mode calibration data could not be reliably used for (relative) flux calibration, and wavelength calibration was also a challenge, especially in the blue arm, where a 2\,\AA \ shift is apparent between arcs taken at the start and the end of both nights.

Standard long-slit, single object reduction procedures were performed in \textsc{iraf} \citep{Tody1986,Tody1993} for spectral extraction. Because a bright standard star was not available, two-dimensional images of the science target were co-added to obtain the optimal trace for each observing night. This trace aperture was then used to extract the individual frames using the \textsc{apall} package. Owing to the lack of reliable flux calibration data, it was decided to calibrate using continuum normalisation. Individual extracted spectra were continuum-normalised by fitting a polynomial over spectral regions that are free of emission or telluric features. A third-order polynomial function was used to normalise individual blue arm spectra, and a fourth order for individual red arm spectra. The overall accuracy of the normalisation was then verified by assessing trailed spectrograms to ensure relative homogeneity between the individual frames. The S/N was assessed for individual spectra in the regions $4500-4600$\,\AA \ and $6600-6700$\,\AA, and found to be $\langle {\rm S/N} \rangle=23\pm2$ for the 2018 July run and $\langle {\rm S/N} \rangle=19\pm2$ for 2019 May.

\begin{table}
\begin{center}
\caption{Published parameters of GD\,356.}
\label{tab:Tab1}
\begin{tabular}{lr}
		
\hline

Parameter					&Value\\

\hline

Spectral Type				&DAHe\\
$V$ (mag)					&15.1\\
$B_{\rm dip}$ (MG)			&$13\pm2$\\
Distance (pc)				&$20.14\pm0.01$\\

$v_{\rm tan}$ (km\,s$^{-1}$)	&$21.47\pm0.02$\\				

$T_{\rm eff}$ (K)			&$7560\pm30$\\
$\log\,g$ (cm\,s$^{-2}$)		&$8.19\pm0.01$\\
Mass (M$_{\odot}$)			&$0.70\pm0.01$\\
Cooling Age (Gyr)			&$1.98\pm0.04$\\

\hline

\end{tabular}
\end{center}
{\em References}: \citet{Ferrario1997a,Bergeron2001,Fontaine2001,Gentile2019}.
\end{table}

\subsection{Spectropolarimetry}\label{Sect_Specpol}

The WHT ISIS instrument was also used to obtain a time series of circular spectropolarimetric measurements on 2019 October 9 over an entire rotational cycle of GD\,356. This strategy allowed a measurement of the polarisation averaged over almost the entire spin cycle, but can also detect variability of the magnetic field due to rotation. Simultaneous exposures were taken through the R600B grating in the blue arm, and the R1200R grating in the red arm, with central wavelengths 4400\,\AA \ and 6500\,\AA \ respectively. The slit width was set to 1.0\,arcsec and spectral resolving powers of $R\approx2350$ and $R\approx7800$ were attained in the blue and the red arm, respectively. Twelve pairs of exposures were obtained with the retarder waveplate at position angles of $\pm45\degr$, thus exchanging the right- and left circularly polarised spectra on the CCDs. This beam-swapping procedure avoids most of the possibly serious systematic errors that arise from single integrations using a fixed waveplate position \citep[e.g.][]{Bagetal09}. An exposure time of 300\,s was adopted for each frame at each position, providing an effective cadence of 600\,s in measuring Stokes $V$ and $I$. For each exposure pair, a global peak S/N of $\approx50$ per \AA \ was achieved in the blue arm, and $\approx40$ per \AA \ in the red arm. 

\begin{figure*}
\includegraphics[width=\linewidth]{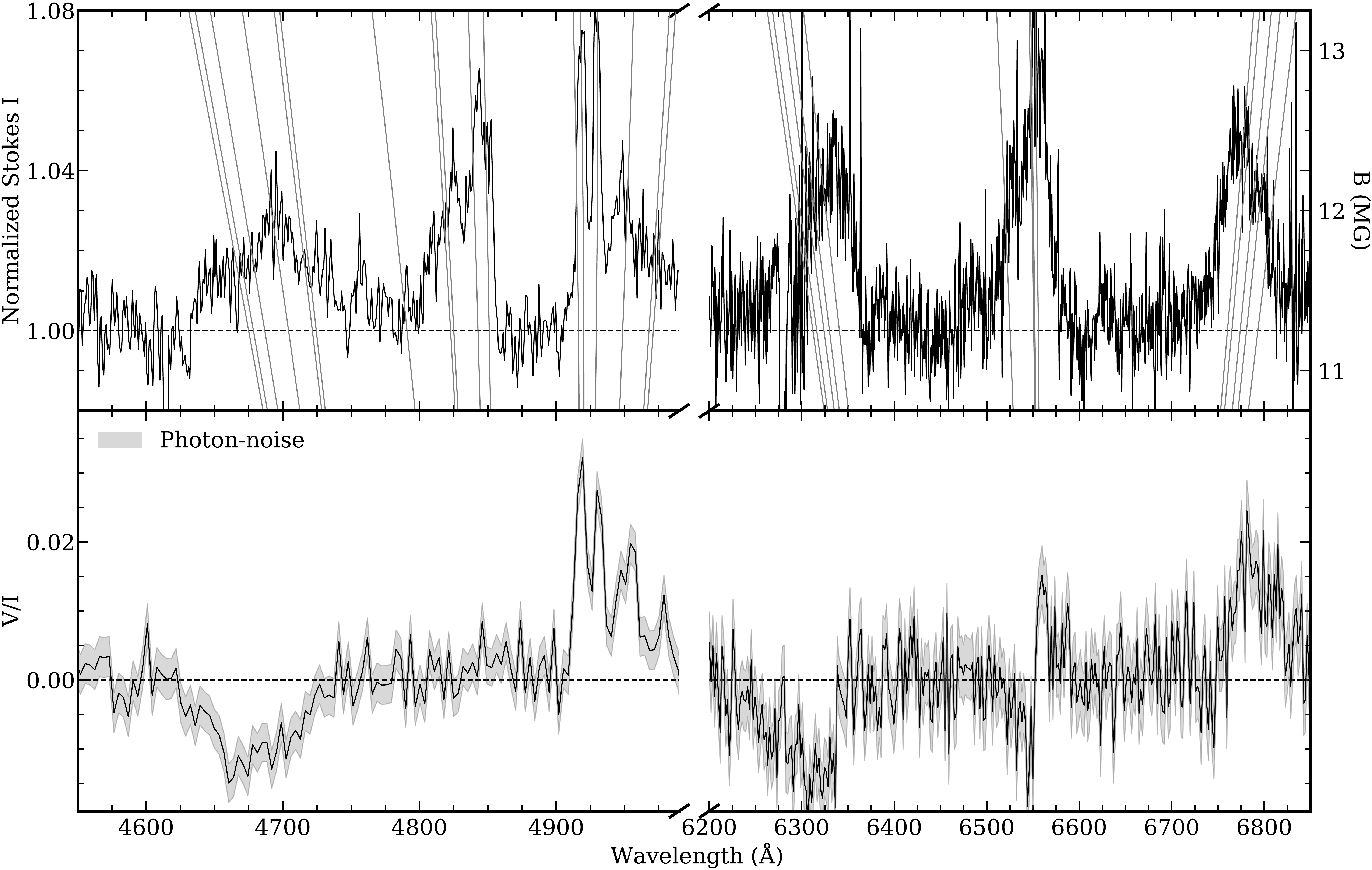}
\vskip 1mm
\caption{\textit{Upper panel}: Integrated 2\,h intensity ($I$) spectrum obtained via spectropolarimetry. Grey lines represent magnetic transitions as a function of field strength \citep{Schimeczek2014}. \textit{Lower panel}: The average $V/I$ spectrum with estimated photon-noise overplotted in grey. The former has the opposite sign to that shown in Fig.~1 of \citet{Ferrario1997a}, which is simply due to different definitions adopted in this work \citep[see Sect.~\ref{Sect_Specpol} and][]{Bagnulo2018}. The spectrum was re-binned for clarity to $2.0$\,\AA \ and $3.5$\,\AA \ in red and blue, respectively.}
\label{fig:Fig2}
\end{figure*}

The details of the data reduction, and a discussion on the conventions adopted for Stokes $V$ are described in \citet{Bagnulo2018}. In agreement with most recent literature on stellar magnetic fields, Stokes $V$ is defined as right-handed circular polarisation minus left-handed circular polarisation (as seen from the observer). Notably, this is opposite the definition adopted by \citet{Ferrario1997a} in their study of GD\,356. The average Stokes $I$ and $V/I$ over a full rotation are shown in (Fig.~\ref{fig:Fig2}).

In order to identify correctly the extent of the Zeeman $\upsigma$ and $\uppi$ components, and because later measurements require an accurately calibrated $I$ spectrum for normalisation of measurements of the polarised (Stokes $V$ or $V/I$) spectra, the instrumental wavelength sensitivity variations from the flux (Stokes $I$) spectra have been removed. This was done through the use of featureless DC white dwarf spectra taken with the same instrumental setting as for the science target. Each spectrum of GD\,356 was normalised to 1.00 near the centre, at 6400\,\AA \ for the red spectra, and at 4500\,\AA \ for the blue spectra. The DC flat field standard was then normalised to 1.00 in the same region, and the slope of its continuum was adjusted by multiplication by a factor of $[1.00 - C((\uplambda - \uplambda_0)/\uplambda_0)]$, where $C$ is a constant adjusted for each spectrum, with a value of around 0.4 for blue spectra flat-fielded with WD\,1055--072 ($T_{\rm eff} = 7155$\,K), and a value around 2.0 for red spectra flat-fielded with WD\,1820+609 ($T_{\rm eff} = 4865$\,K). This fitting procedure led to virtually identical spectra over several hundred \AA \ of spectral data exterior to the extended Zeeman components of GD\,356. The process was then completed by dividing the spectrum of GD\,356 by the matched spectrum of the DC white dwarf, resulting in flux spectra accurately normalised to unity except within the Zeeman components.

\begin{table}
\begin{center}
\caption{Ground-based, time-series observations obtained for this study.}
\label{tab:Tab2}

\begin{tabular}{@{}lcccrr@{}}

\hline

Telescope	/	&Obs	&Obs 		&Duration	&$t_{\rm exp}$	&$n_{\rm exp}$\\
Instrument	&Type	&Date		&(h)		&(s)			&\\

\hline

WHT / ISIS	&Spec	&2018 Jul 29 	&4.4		&120		&113\\
WHT	 / ISIS	&Spec	&2019 May 14 	&4.6		&120		&120\\
WHT	 / ISIS	&Spol	&2019 Oct 09	&2.0		&300		&24\\
LT / RISE		&Phot	&2020 Jul 09	&2.0		&6		&1168\\
LT / RISE		&Phot	&2020 Jul 10	&2.0		&6		&1168\\
LT / RISE		&Phot	&2020 Jul 14	&2.0		&6		&1168\\
LT / RISE		&Phot	&2020 Jul 17	&2.0		&6		&691\\
LT / RISE		&Phot	&2020 Jul 26	&2.0		&6		&1168\\
LT / RISE		&Phot	&2020 Aug 08	&2.0		&6		&1168\\
LT / RISE		&Phot	&2020 Aug 16	&2.0		&6		&1168\\
PTO / PRISM  	&Phot  	&2020 Aug 15	&3.3 		&10 		&784\\

\hline

\end{tabular}
\end{center}
\end{table}

\subsection{{\em TESS} observations}

GD\,356 was observed by \textit{TESS} in Sectors 16, 23, 24, 25, and 26 under designation TIC\,274239484 (\textit{TESS} Mag = 14.8\,mag; \citealt{Stassun2018}). Data from Sector 16 were observed from 2019 September 12 until 2019 October 6. Data from Sector 23 onward span dates from 2020 March 19 until 2020 July 4. \textit{TESS} utilises a red-optical bandpass covering roughly $6000-10\,000$\,\AA, and each CCD has a $4096\times4096$\,pixel$^2$ on-sky area, which are read out at 2\,s intervals. The on-board computer then produces $11\times11$\,pixel$^2$ postage stamps centred at the target, with flux averaged over 120\,s. After a quality-flag cut, a total of 113.7\,d of GD\,356 observations at 2\,min cadence were retained for photometric analysis. Only points with the nominal quality flag were analysed (94.4\,per cent of the total recordings).

Data products from the outlined sectors were accessed from the \textit{TESS} Mikulski Archive for Space Telescopes pages\footnote{\url{https://mast.stsci.edu/}}. Light curves were extracted from the 2\,min target pixel files by the \textit{TESS} Science Processing Operations Center pipeline \citep{Jenkins2016}. Pre-search Data Conditioning Simple Aperture Photometry (PDCSAP) light curves were used, as these have systematic errors removed, including error sources from the telescope and the spacecraft \citep{Smith2012,Stumpe2012}. Flux values were divided by the mean for each sector and centred at zero. Aperture masks identified by the pipeline differed between the sectors, and to investigate possible effects of this, the \textsc{lightkurve} package was used to extract the data with a consistent aperture mask. No significant differences in the resulting light curves were found, and therefore, pipeline defined apertures were used throughout for reliability and reproducibility provided by the PDCSAP. 

{\em Gaia} DR2 reveals a source that is 12.8\,arcsec distant from GD\,356 and fainter by $\Delta G=3.2$\,mag, and $\Delta R_{\rm P}=2.4$\,mag \citep{Gaia2018}, thus on the order of ten times fainter in the region of the {\em TESS} bandpass. There appears to be no significant contamination from this source in the extracted data. From the {\em TESS} pipeline, an average ratio of target flux to total flux (\textsc{crowdsap}) in the aperture is 0.87 over five sectors. This is accounted for in the PDCSAP data. 

\subsection{The Liverpool Telescope}

Ground-based photometry was acquired on the fully robotic 2.0\,m Liverpool Telescope (LT) at Roque de los Muchachos, La Palma. The fast-readout CCD imager, Rapid Imaging Search for Exoplanets (RISE; \citealt{RISE2008}) instrument was used on 2020 July 9, 10, 14, 17, and 26, as well as on 2020 August 8 and 16. The imaging data were taken with the in-house $V+R$ filter, constructed from a 3\,mm Schott OG515 bonded to 2\,mm Schott KG3 filter, and with a central wavelength of approximately 5900\,\AA. The data were acquired in the $2\times2$ binning configuration, corresponding to 1.17\,arcsec$^2$ per pixel$^2$ on an E2V CCD47-20 frame-transfer CCD with a $1024\times1024$\,pixel$^2$ light-sensitive region. All but one observational group covered the 1.93\,h rotational period of GD\,356, including all overheads, with 6\,s individual exposures, yielding a total of 1168 frames per group. The remaining observational visit on 2020 July 17 was affected by the calima (dust) resulting in an early dome closure. The data were automatically reduced through the RISE pipeline which performs bias subtraction, removes a scaled dark frame, and generates the flat-field correction. Overall, a total of seven sets of data were obtained, six of which cover the full photometric cycle of GD\,356 at a higher cadence, a finer spatial scale, and higher S/N compared to \textit{TESS}.

Circular aperture photometry was performed in \textsc{AstroImageJ} \citep{Collins2017} to obtain relative fluxes comprised of the net integrated counts of the target divided by the total integrated counts of all comparison stars. Three comparison stars were used for differential photometry, where the faintest was approximately three times brighter within the photometric aperture than GD\,356. These comparison sources appear non-varying, with S/N always greater than the target, yet below the level where the CCD has a non-linear response. Aperture size was chosen empirically and individually for each visit based on uncertainty minimisation of the resulting amplitude of the photometric variability, where a typical aperture radius was 4\,arcsec or roughly twice the typical seeing ($2-2.5$\,arcsec). Finally, a barycentric correction was applied to the mid-points of the exposure time stamps to allow for a comparison with other photometric studies.

\subsection{The Perkins Telescope Observatory}

Additional ground-based photometry was acquired on the 1.8\,m Perkins Telescope Observatory (PTO) on Anderson Mesa outside of Flagstaff, Arizona, on 2020 August 15, using the Perkins Re-Imaging SysteM (PRISM; \citealt{Prism2004}). Time-series photometry was performed in the SDSS $g$ filter with no binning, with each exposure lasting 10\,s. The readout times are typically slow (8.0\,s) for PRISM, but they were reduced by using a $410\times410$\,pixel$^2$ subarray of the CCD, reduced from its full $2048\times2048$\,pixel$^2$. This strategy yielded 784 frames over 196\,min, including overheads. The weather was clear, and the seeing ranged from $2.1-4.3$\,arcsec. The data were reduced with custom code written in {\sc python} that corrects the data for bias, performs the flat field, and executes aperture photometry using \textsc{Photutils} version 0.7.2 \citep{Photutils2019}. Frame times were corrected to barycentric times using \textsc{AstroPy} \citep{astropy1, astropy2}. At the end of the run, a roughly 3\,s offset was found in the data-acquisition computer time compared to a Network Time Protocol time server, which was corrected by adding 3.0\,s to each time in the light curve. Given these imprecise time stamps, 3.0\,s was also added to the estimated phase uncertainty.

\begin{figure}
\includegraphics[width=\columnwidth]{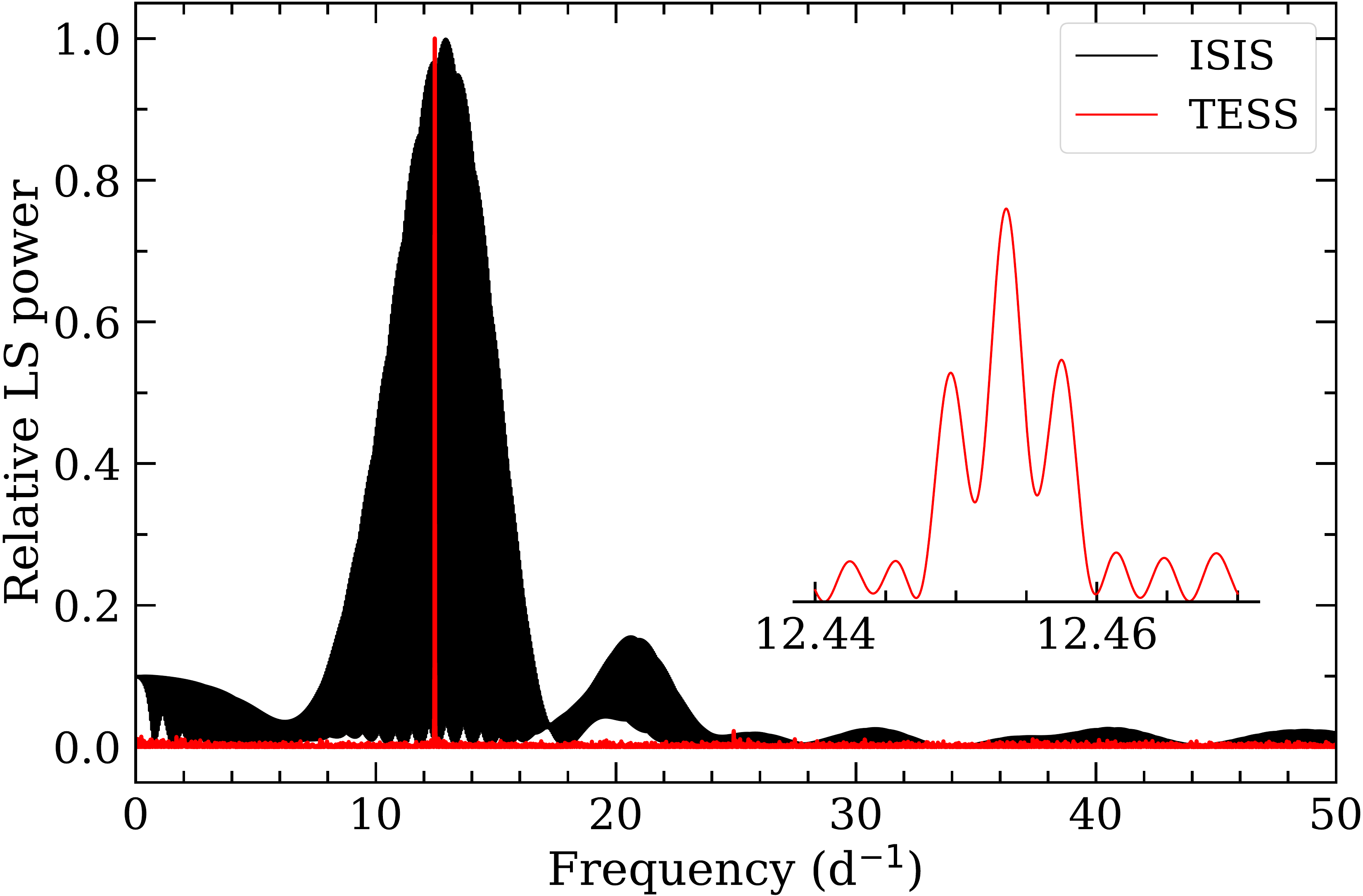}
\vskip 1mm
\caption{Lomb-Scargle (LS) periodogram of ISIS relative equivalent widths ($W_\uplambda$, black) and \textit{TESS} photometry (red).  The \textit{TESS} data show a clear peak at 12.45\,d$^{-1}$ (zoomed panel) corresponding to the known period, and a less obvious but significant first harmonic of this frequency is also present at approximately 25\,d$^{-1}$.}
\label{fig:Fig3}
\end{figure}

\begin{figure}
\includegraphics[width=\columnwidth]{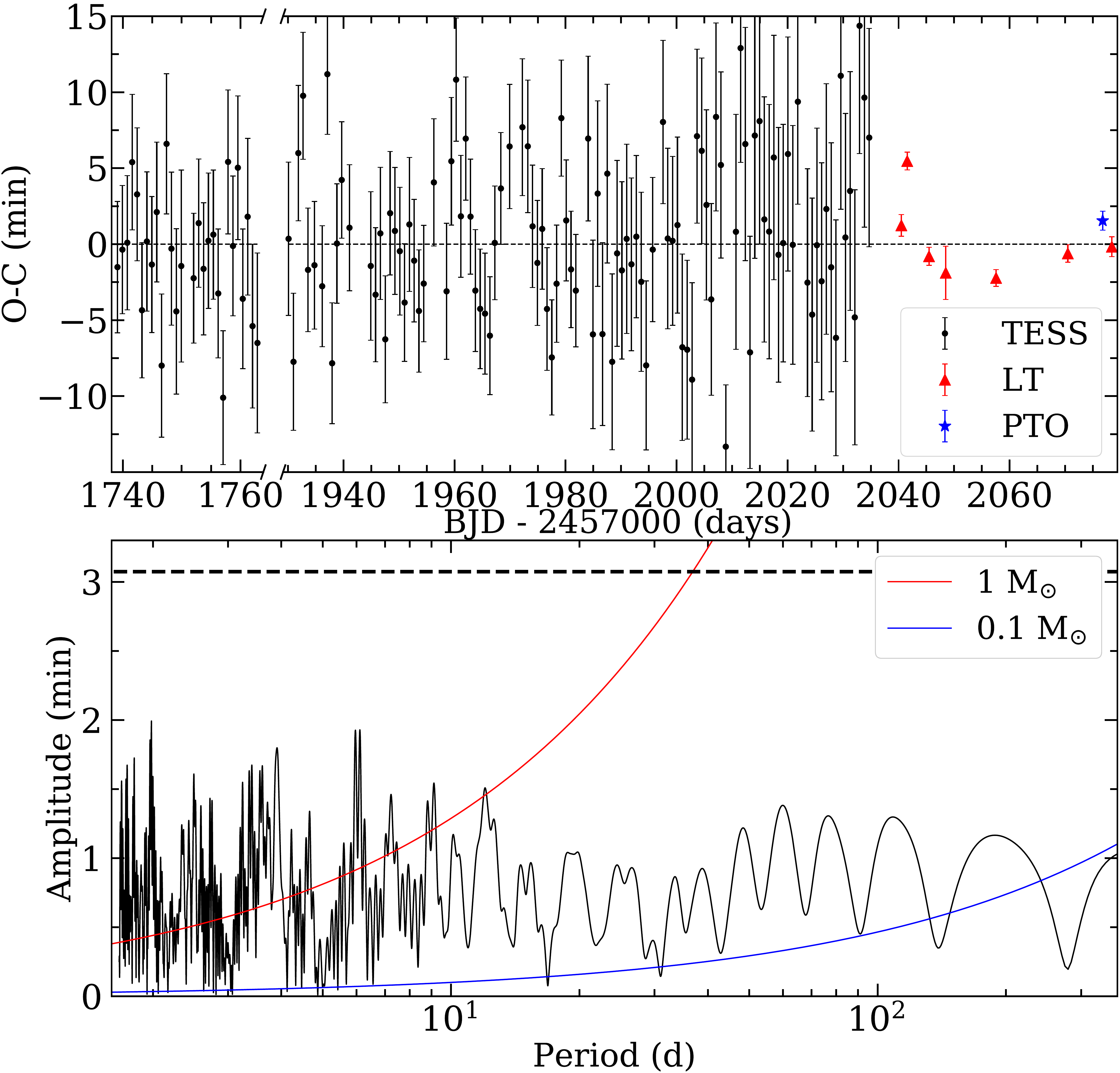}
\vskip 1mm
\caption{\textit{Upper panel}: $O-C$ diagram of \textit{TESS} sectors 16, 23, 24, 25, and 26, and ground-based LT and PTO observations showing the arrival time of the rotational signal assuming a constant spin period. Each \textit{TESS} data point is a time slice of approximately 20\,h or roughly 10.5 rotational cycles. Modest discrepancies are seen between the expected and observed arrivals, in both space- and ground-based observations. \textit{Lower panel}: LS periodogram of the $O-C$ data plotted above, where the dotted line just above 3\,min in amplitude is an empirical estimate of the 1\,per cent false-alarm probability. Although the photometric data deviate somewhat from a constant in $O-C$, there is no statistically significant signal in the periodogram. Moreover, the expected amplitude of a variation induced by a planetary-sized companion would be much smaller than can be constrained by these data. This point is demonstrated by the blue and red lines which display the expected amplitudes from a 0.1 and 1.0\,M$_\odot$ companion for a range of orbital periods.}
\label{fig:Fig4}
\end{figure}

\section{Analysis and Results}\label{sec:res}

First the photometric data are presented, as the known -- now updated and highly precise -- photometric period sets a foundation against which all observed variability may be compared. Owing to extensive \textit{TESS} observations, the photometric period is sufficiently constrained to allow for comparisons of epochs multiple years apart. To asses the origins of the broad-band photometric variability in GD\,356, the semi-amplitude of the light curves in several bandpasses are compared with that expected from changes in the emission lines alone. Following these results, both the time-series spectroscopy and spectropolarimetry are discussed. 

\subsection{Photometric Variability}


To establish a physical context for any and all spectroscopic or polarimetric variability, the \textit{TESS} observations are examined. The star has unparalleled temporal coverage from a total of five sectors, four of which are consecutive, and together with the 2\,min cadence provide an excellent opportunity to investigate the photometric variability in detail. The \textsc{AstroPy} implementation\footnote{The \textsc{AstroPy} Lomb-Scargle documentation available at: \url{https://docs.astropy.org/en/stable/timeseries/lombscargle.html}} of the Lomb-Scargle (LS; \citealt{Lomb1976,Scargle1982}) periodogram for the pipeline-processed light curves shows a strong signal at $1.92717 \pm 0.00001$\,h (Fig.~\ref{fig:Fig3}), and is the most precise determination of the photometric period to date. The period determined from these data are within $2\upsigma$ of the value established from $V$-band observations almost 20 years prior \citep{Brinkworth2004}. The observed semi-amplitude of the variation is $0.64\pm0.02$\,per cent in the {\em TESS} filter, which fully covers all three H\,$\upalpha$ emission components, whereas the $V$-band observations had an semi-amplitude of 0.2\,per cent and this filter covers only a fraction of the H\,$\upbeta$ emission. Despite the relatively small uncertainty in the period, the direct comparison of $T_0$ from {\em TESS} and \citet{Brinkworth2004} is not possible.

The unipolar inductor predicts that the hot emission region is the result of a current loop connecting the stellar surface with an orbiting and conducting planetary body. It is thus expected that such a signal would be present in photometric data, assuming the spot location is (even partly) modulated by its Keplerian orbit. First, the only significant signals seen in the photometric data analysed here are the rotational frequency and its first harmonic. Second, an observed (O) minus calculated (C) arrival time diagram is shown in Fig.~\ref{fig:Fig4}, where both {\em TESS} and ground-based observations are plotted. Despite the profound temporal coverage of {\em TESS}, and higher accuracy of the ground-based arrival times, there are no clear trends or periodic signals in the $O-C$ data -- certainly nothing above the false alarm probability \citep{Greiss2014}. Therefore, evidence of a secondary modulation in the photometric period has not been detected in the extensive \textit{TESS} or ground-based data, nor has the period appeared to change in more than 15 years.

\begin{table}
\begin{center}
\caption{Re-constructed spectral energy distribution of GD\,356 convolved with the various light curve filters.}
\label{tab:Tab3}

\begin{tabular}{@{}llccccr@{}}

\hline

Facility    		&Filter		&$\uplambda_{\rm eff}$	&$f_{\rm em}$	&$A_{\rm W}$   &$A_0$	&Year\\
            		&       		&(\AA)                  		&(\%)		&(\%)		&(\%)	&\\

\hline
PTO		    	&$g$	    		&4750				&0.69		&0.13 		&1.28	&2020\\
JKT		    	&$V$        		&5510				&0.11		&0.02 		&0.20	&2002--2003\\
LT		   	&$V$\!+$R$      &6170				&0.27		&0.05 		&0.83	&2020\\
{\em TESS}	&{\em TESS}	&7900				&0.18		&0.03 		&0.65	&2019--2020\\

\hline

\end{tabular}
\end{center}
{\em Note}. The third column is the flux-weighted, average wavelength, and the fourth column gives the fractional contribution of the emission lines relative to the continuum. The fifth column is the expected semi-amplitude $A_{\rm W}=f_{\rm em}\times0.19$ from only the observed variation in $W_\uplambda$ (whose semi-amplitude is 0.19 for all emission lines). The sixth column is the observed light curve semi-amplitude, and is an order of magnitude larger than $A_{\rm W}$ in all cases.
\end{table}

The lack of additional frequencies does not support a unipolar inductor with planetary orbits that are well-sampled by the {\em TESS} light curves (on the order of hours to weeks). However, the $O-C$ observations are hardly constraining in terms of orbital modulation. A maximum amplitude in the phase variation can be calculated for a potential companion \citep{Hermes2018}. As shown in Fig.~\ref{fig:Fig4}, companions of sensible masses cannot be ruled out from these data, e.g.\ 1.0\,M$_\odot$ object on a 30 day orbit can be dismissed. Clearly, the existing $3-8\,\rm\upmu m$ \textit{Spitzer} constraint of 12\,M$_{\text{Jup}}$ \citep{Wickramasinghe2010} is far more informative. Potentially, future \textit{TESS} observations might prove useful in a prospective $O-C$ interpretation.  However, there does appear to be at least one point based on LT observations that is more than $3\upsigma$ from zero in Fig.~\ref{fig:Fig4}, and may indicate some modest surface feature evolution.
But from these analyses, it is the lack of additional frequencies that fails to indicate any significant emission region modulation by a conducting planetary companion, as in the unipolar inductor model.  

Emission line variability seen in the ISIS spectra cannot be solely due to continuum variation, because the relative semi-amplitude of equivalent width variation $0.19\pm 0.01$ (Fig.\,\ref{fig:Fig5}) is inconsistent with photometric semi-amplitudes (below 0.015 for PTO, LT and \textit{TESS}). Thus, the emission must be contributing to the variability, albeit the anti-phase relation between equivalent widths and photometry. To quantify the contribution of emission a synthetic spectrum was stitched up that includes H\,$\upalpha$ to H$\upgamma$ emission. This spectrum was then continuum normalised and the most prominent telluric features were removed. Next the spectrum was multiplied by a normalised black body model at $T=8000$\,K. The processed spectrum was then multiplied by a transmission function of the photometric filters used in this investigation and the previous photometric study \citep{Brinkworth2004}. Results are summarised in Table \ref{tab:Tab3}. It appears that the dominant source of the photometric variability must be due to continuum variation, since the emission line variability is approximately an order of magnitude below the observed values. The continuum variation hypothesis is further supported by the fact that the variation is larger at shorter wavelengths, which is consistent with a blackbody at 8000\,K.

\subsection{Spectroscopic Variability}

The ISIS long-slit (non-polarimetric) spectra exhibit Balmer emission features at all rotational phases (Fig.~\ref{fig:Fig1}). The known H\,$\upalpha$ and H\,$\upbeta$ features are split into Zeeman triplets, where the central $\uppi$ components exhibit a weak displacement relative to the zero-field position, and the side components $\upsigma_{\rm b}$ (blueshifted) and $\upsigma_{\rm r}$ (redshifted) are displaced approximately equally from the undisturbed spectroscopic location and carry opposite polarisation states Fig.~\ref{fig:Fig2}. There is also a single, weak emission feature observed at H$\upgamma$ (not shown), where other Zeeman components are not detected, even in a co-added spectrum of an entire night.

\begin{figure}
\includegraphics[width=\columnwidth]{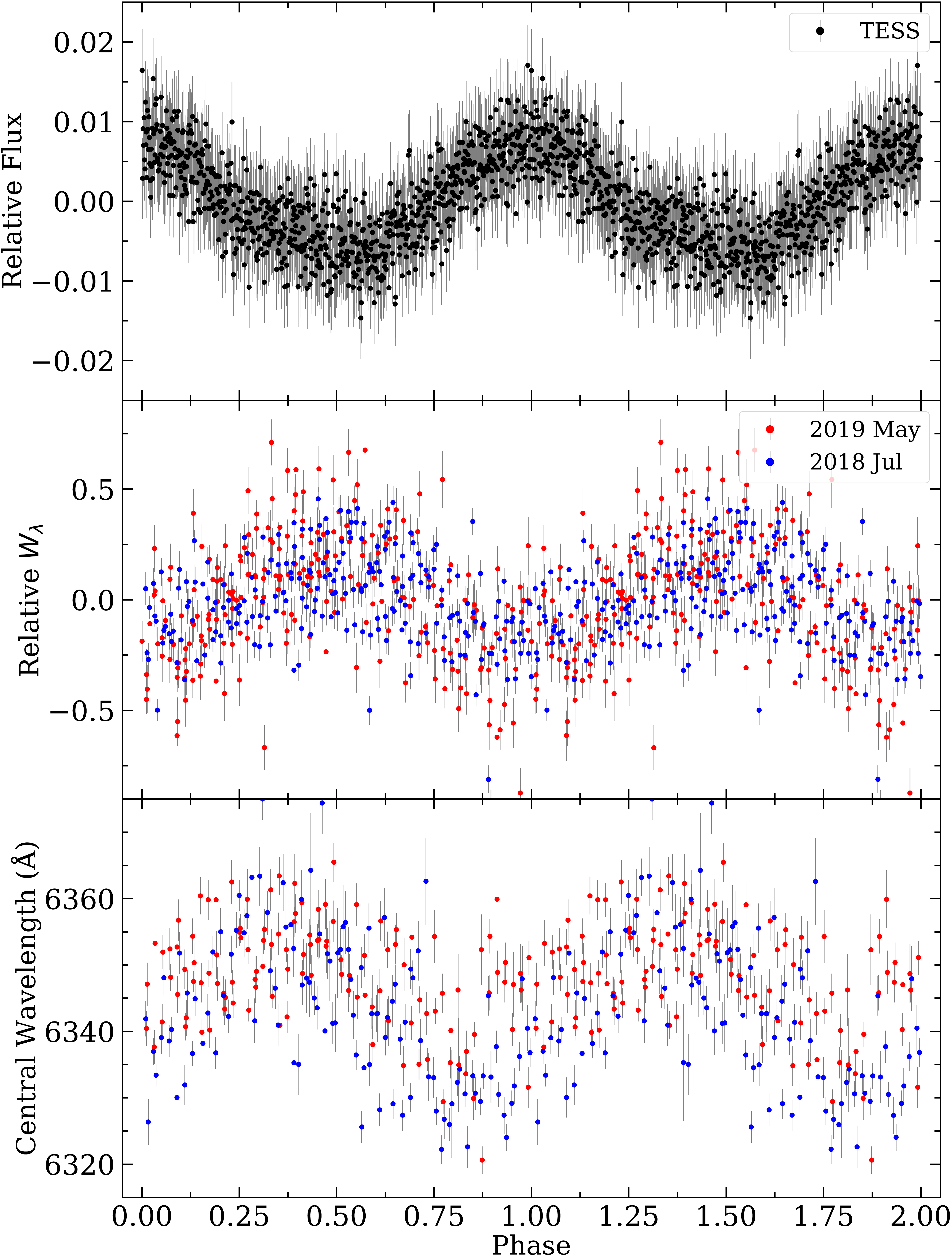}
\vskip 1mm
\caption{\textit{Upper panel}: All five available \textit{TESS} sectors combined and folded on the 1.92717\,h period, where each point is an average of 100 phase-consecutive recordings. \textit{Middle panel}: Relative $W_\uplambda$ estimates from ISIS spectra coloured by the observing run. \textit{Lower panel}: Variation in the central wavelengths of the $\upsigma_{\rm b}$ component of H\,$\upalpha$. A change in the central wavelength of $-22$\,\AA\: corresponds to a change in the average magnetic field of $+1.0$\,MG. Phase $=0$ corresponds to photometric maximum in all three panels.}
\label{fig:Fig5}
\end{figure}

The equivalent widths ($W_\uplambda$; \citealt{Vollmann2006}) are measured to investigate the emission activity of GD\,356, using determinations for each H\,$\upalpha$ and H\,$\upbeta$ Zeeman component, corresponding to six total $W_\uplambda$ values per time-series spectra. The individual component $W_\uplambda$ are divided by the temporal mean, and all six spectral components are averaged into a single $\langle W_\uplambda \rangle$ per time-series exposure, which is then passed to the periodogram algorithm. 

Computing the LS periodogram for the $W_\uplambda$-time recordings shows a strong signal near the known photometric period (Fig.~\ref{fig:Fig3}). Furthermore, a repeating, sinusoidal pattern is evident in $W_\uplambda$ for each of the six individual emission components, where the clarity of the pattern varies depending on the particular feature. To better quantify these time-dependant changes for individual features, a goodness of fit statistic is estimated for each emission component using the spin frequency, and confirms the same variability is present in all emission feature components, and above the typical noise level in each case. Variation is conspicuous in the $\uppi$ components of H\,$\upalpha$ and H\,$\upbeta$, as well as the $\upsigma_{\rm r}$ components of H\,$\upbeta$. The behaviour of $W_\uplambda$ values during the 2018 July and 2019 May observing runs are entirely consistent (Fig.~\ref{fig:Fig5}).

Interestingly, when the $\langle W_\uplambda \rangle$ of the emission is phase-folded on the photometric period, it displays a $180\degr$ phase shift; the photometric maximum occurs when $\langle W_\uplambda \rangle$ is minimum and vice versa (Fig.~\ref{fig:Fig5}). To quantify the phase shift, we fit the respective sinusoids to measure a relative phase and its uncertainty. A normalised phase discrepancy of $0.50\pm0.01$ is found between the \textit{TESS} photometry and ISIS equivalent widths. This anti-phase behaviour between photometric flux and emission line strength is corroborated by the LT photometry, whose phase is consistent with that of {\em TESS} (Fig.~\ref{fig:Fig4}).

From the visual inspection of co-added spectra that consist of five consecutive images (Fig,~\ref{fig:Fig1}), there are some notable morphological changes in the central emission features over the spin period. These changes in the shape of the emission lines are most notable in the $\uppi$ component of H\,$\upbeta$, and possibly in the $\uppi$ component of H\,$\upalpha$, and appear periodic with rotation. Because each emission component is a combination of contributions from multiple magnetic-atomic transitions, the morphological modulation is likely due to periodic displacement of individual transitions in response to a changing view of the local magnetic field.  It should be noted that this morphological variability alone is unlikely to be the primary source of $W_\uplambda$ modulation, because morphological changes are not detected in other emission components with clear $W_\uplambda$ periodicity. Thus, the observed variability in the equivalent widths is likely due to a variation in the emission strength rather than magnetically induced changes in the emission profile morphology.

A variation in the mean wavelength of the emission profiles is also apparent. Each emission feature has been fitted with a Gaussian by minimising the difference between the fit and the profile. The parameter that corresponded to the mean of the Gaussian fit is then used as a proxy for the centre of the emission, where the nature of the fitting method results in a more stable fit to broader emission components. Partly for this reason, the $\upsigma_{\rm b}$ component of H\,$\upalpha$ exhibits the clearest periodic variation in wavelength and is shown in the bottom panel of Fig.~\ref{fig:Fig5}. Interestingly, while the variability is unsurprisingly periodic over a rotational cycle, there is a noticeable, normalised phase shift of $-0.13$ relative to the $W_\uplambda$ maximum ($+0.37$ relative to the photometric maximum).

These central wavelengths of the emission profiles act as an excellent indicator of the average magnetic field in the emission region, where the larger blueshift in the $\upsigma_{\rm b}$ component of H\,$\upalpha$ corresponds to a larger magnitude of the surface-averaged magnetic field. Hence, the higher field values approximately correspond to higher photometric amplitudes, although the $0.13$ relative phase shift indicates that this correlation is likely accidental. This fact also suggests that the position of the emitting region does not favour the extremes of the magnetic field, at least in a context of the surface-averaged field. The lack of a direct relationship between the average magnetic field and photometry is potentially further supported by the spectroscopy of another DAHe star \citep{Reding2020}, where the maximum field value roughly corresponds to the photometric minimum.

\begin{figure}
\includegraphics[width=\columnwidth]{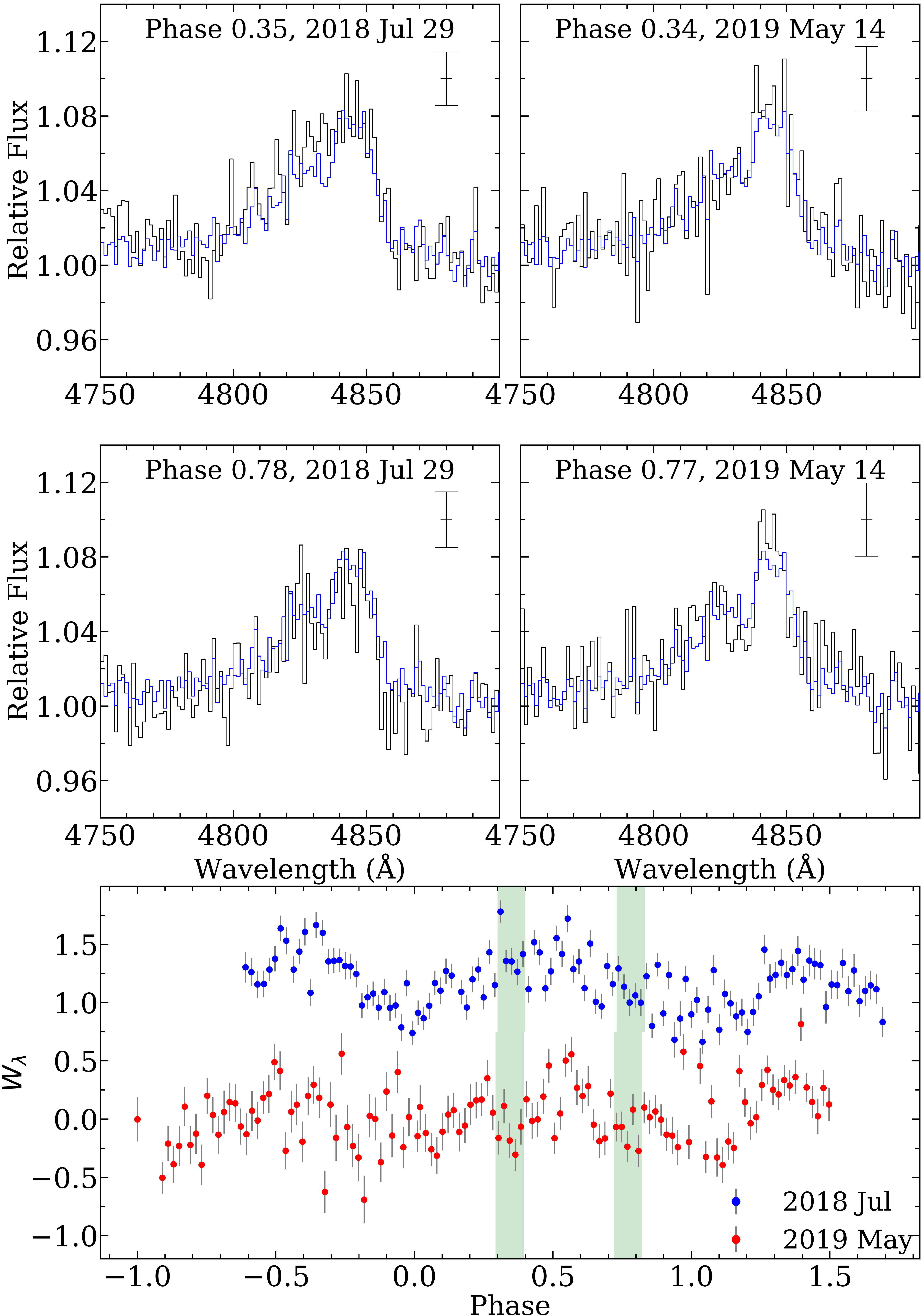}
\vskip 1mm
\caption{\textit{Top four panels}:  Potential morphological dissimilarity in the $\uppi$ component of H\,$\upbeta$ at approximately the same phase of rotation observed on different observing runs separated by ten months. Each panel represents 10\,min of spectroscopy, where the spectra are re-binned for visual clarity. A blue line represents the average line profile across the entirety of the two observing runs for reference. Finally, one standard deviation is given as a black error bar for each panel. \textit{Lower two panels}:  Normalised $W_\uplambda$ for the region spanning $4822-4844$\,\AA \ only, where the green shading highlights the rotation phase corresponding to the top four panels, and the 2019 run is offset vertically downward for clarity.}
\label{fig:Fig6}
\end{figure}

\subsection{Possible Emission Feature Evolution}\label{Same_phase_diff}

While it is clear that the $\uppi$ component of H\,$\upbeta$ varies on the same timescale as the photometric flux, it is possible that the two ISIS data sets exhibit morphological dissimilarities in this feature at the same rotational phase (Fig.~\ref{fig:Fig6}). The two observing runs are separated by nearly ten months, and in particular the 2019 May data display a dip in the $\uppi$ component for a larger part of the cycle, for example. Initially, these dissimilarities were considered from the viewpoint of the unipolar inductor model. To investigate this possibility further,  $W_\uplambda$ are measured for the narrow region $4822-4844$\,\AA \ where these superficial differences can be seen from visual inspection. These area estimates capture potential differences that are clearest near normalised phase 0.35 and 0.78 highlighted in the lower panel of the figure. There is a similar, modest difference between the 2018 July and 2019 May data $W_\uplambda$ of all three H\,$\upbeta$ components (not shown), at the same phase. The data do not permit a more robust analysis and the reality of these possible feature evolution over many months is intriguing but uncertain.

Alternatively, assuming the variations presented in the $O-C$ diagram (Fig.~\ref{fig:Fig4}) are physical, then the morphological differences in the emission at the same rotational phase could be linked to small deviations in the photometric period. Given the exact anti-phase relationship between the photometric and emission intensity, the two quantities could be closely related. Therefore, judging by the fact that some inconsistencies are seen with the fixed period sinusoidal model in photometry, one might expect to see the same behaviour in the emission. Given that the emission profiles were analysed from co-added spectra with combined exposure of ten minutes, then a small alteration in phase of a few minutes could produce the observed behaviour and would be consistent with the $O-C$ photometric analysis. This would indicate that the features are evolving, albeit modestly at most, on a timescale of days rather than months.

\subsection{Spectropolarimetric Measurements and Variability}

\begin{figure}
\includegraphics[width=\columnwidth]{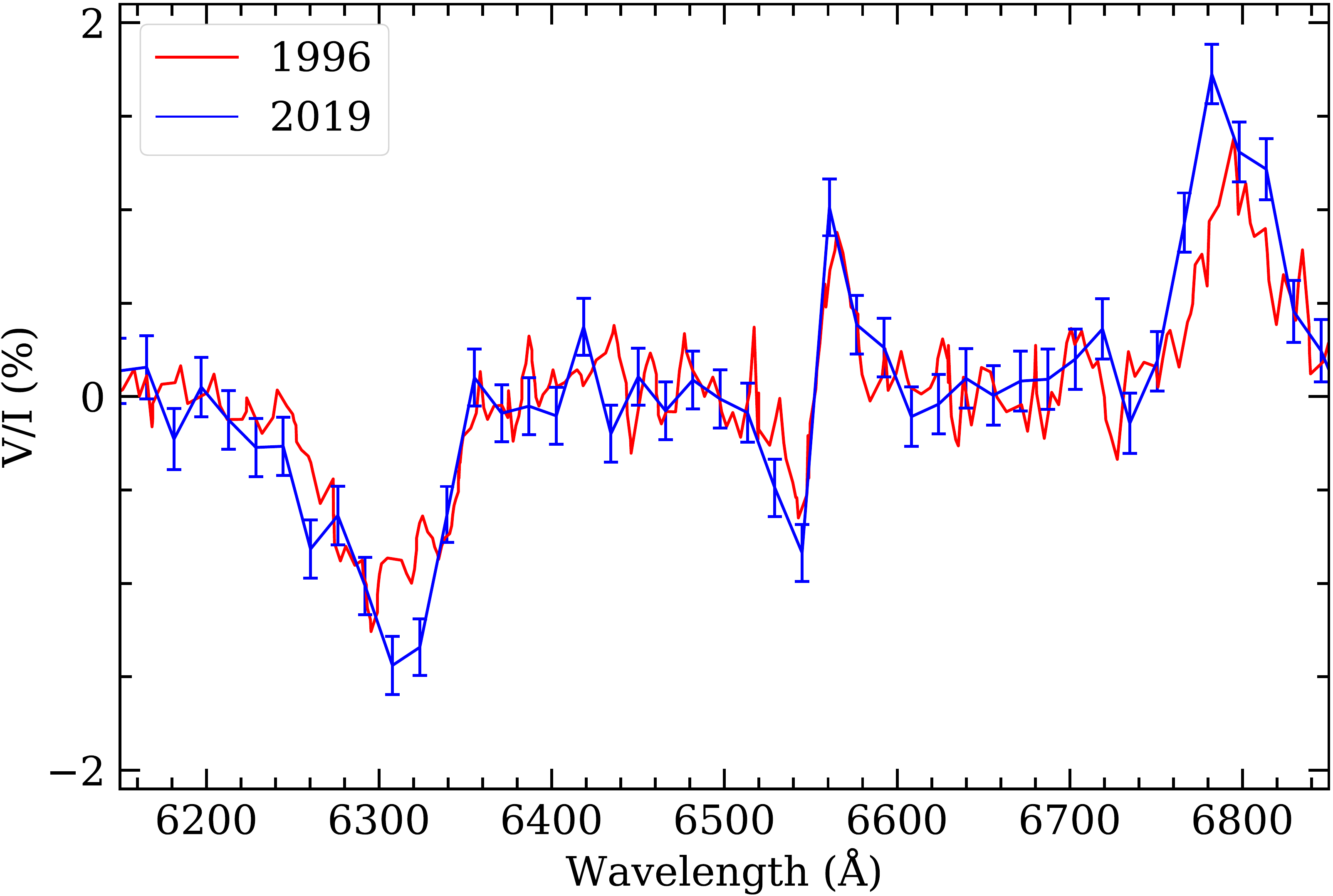}
\vskip 1mm
\caption{Comparison of circular spectropolarimetric data digitalised from Fig~4 of \citet{Ferrario1997a} and those presented in this work, re-binned at 16\,\AA. The sign of the data from \citet{Ferrario1997a} has been changed to make it consistent with the convention used here.}
\label{fig:Fig7}
\end{figure}

The spectropolarimetric observations of GD\,356 provide simultaneous flux and circular polarisation spectra covering one full rotation of GD\,356 with 10\,min time resolution. The polarised spectra cover H\,$\upalpha$ in a window extending from 6200 to 6900\,\AA, and H\,$\upbeta$ and H$\gamma$ in a window extending from 4000 to 5100\,\AA. A first qualitative assessment of these data compares the new circular polarisation data, integrated through the stellar rotation, with the spectropolarimetry of \citet{Ferrario1997a}. This comparison is shown in Fig.~\ref{fig:Fig7}, where the 1997 data (not available in tabular form) have been digitised from Fig.~4 of \citet{Ferrario1997a}. The two $V/I$ spectra are sufficiently similar that they do not offer strong evidence for long-term secular variation of the average polarisation of GD\,356 over a 25\,yr time base.

The \bs\ values are evaluated by computing the mean wavelength of each of the two $\upsigma$ components for both H\,$\upalpha$ and H\,$\upbeta$, based on the excess flux above the continuum in the flat-fielded spectra. The mean longitudinal magnetic field \bz\ is determined following the methods developed for magnetic Ap stars (e.g.\ \citealt{Babcock1947,Mathys1989}), that is, by measuring the wavelength shift of the entire emission feature as observed in right and left circularly polarised light from the unshifted position. It is easily shown that this shift is given by the first wavelength moment of Stokes $V$ around the mean wavelength $\uplambda_0$:

\begin{equation}
    \Delta \uplambda_z = \frac{\int V (\uplambda - \uplambda_0)\, {\rm d}\uplambda} 
       {\int I\, {\rm d}\uplambda}
       \label{Eqn1}
\end{equation}

\smallskip
\noindent
Evaluation of this quantity and its uncertainty are discussed by \citet{Landstreet2015}. 

It is known from previous studies and from Fig.~\ref{fig:Fig2} that the mean field modulus is about 11\,MG \citep{Greenstein1985,Ferrario1997a}. Using the computed wavelength positions of the various $\upsigma$ components as computed and shown in Fig.~\ref{fig:Fig2} \citep{Schimeczek2014}, it is found that at this field strength, the mean separation $\Delta \uplambda$ of the two $\upsigma$ components of H\,$\upalpha$ is given accurately by $\Delta \uplambda = 39.98 B$ where $\Delta \uplambda$ is measured in \AA \ and $B$ in MG (this is almost exactly the conversion constant found for weak fields). The \bs\ uncertainty is evaluated from the uncertainty in the computed mean position of the $\upsigma$ components. It is noteworthy that because the emission features are shallow, the main obstacle to correctly evaluating \bz\ is the flat-fielding of the $I$ spectrum required for computation of the denominator in Eq.~(\ref{Eqn1}).

The resulting \bs\ and \bz\ values are listed in Table\,\ref{tab:Tab4}, in which the first line refers to the integrated 2\,h spectrum. The remaining entries report measurements of the 10\,min observations in order throughout the entire observation, where the mean time is given for each set of exposures. The table also reports other quantities measured from the combined $I$ and $V$ polarised spectra: the mean shift $\Delta \uplambda_Z$ of the right or left circularly polarised spectra from their mean value, the \bz\ values derived from $\Delta \uplambda_Z$, and the equivalent width of each $\upsigma$ component, $W_{\rm \uplambda b}$ and $W_{\rm \uplambda r}$. 
In addition to the mean field modulus and mean longitudinal field, an interesting quantity is the range $\delta B$ of $B$ values present in the emission region. This range may be estimated from the excess width of each $\upsigma$ component over the width -- due to line broadening and splitting of components by the quadratic Zeeman effect -- that the component would have in a uniform $B$ field. Looking at the distribution of the individual $\upsigma$ components of the Balmer lines in a 11.5\,MG field (Fig.~\ref{fig:Fig2}), it is apparent that individual components are more widely separated (relative to the scale of the $\uppi$ -- $\upsigma$ separation) in H\,$\upbeta$ than in H\,$\upalpha$, and thus the most robust $\delta B$ estimates will result from the analysis of H\,$\upalpha$. The range $\delta B$ of field values are estimated using the dispersion of the intensity with wavelength of each $\upsigma$ feature within H\,$\upalpha$, using the same conversion between wavelength and field strength as above, for the full 2\,h averaged spectrum.

Before interpreting the broadening of the $\upsigma$ components in terms of a variation of the magnetic field over the stellar surface, alternative broadening mechanisms of the $\upsigma$ components should be considered. The intrinsic width of a single emission $\upsigma$ component may be estimated from the widths of the two individual stationary red $\upsigma$ components of H\,$\upbeta$ at 4920 and 4930\,\AA, to a value of the order of 5\,\AA. Intrinsic broadening is thus unimportant. Another consideration is the dispersion in wavelength of the individual Zeeman components at 11.5\,MG. Fortunately, the strongest of the five components in each $\upsigma$ pattern are in the middle, while the two outermost components are weakest. 

For H\,$\upalpha$, it is found that at 11\,MG, the dispersion of the theoretically computed wavelengths of the five $\upsigma$ components, weighted by line strength, is only about 7\,\AA. In contrast, the dispersions of the blue and red $\upsigma$ features of H\,$\upalpha$ observed in the intensity spectrum of GD\,356 are about 26 and 19\,\AA, respectively. This suggests that the observed broadening of each feature is due to the spread in the magnetic field strengths over the emission area. Because of the curvature present in the wavelength vs. magnetic field strength for the Zeeman components at 11.5\,MG (see Fig.~\ref{fig:Fig2}), the slope of the conversion between wavelength broadening to field variation is $-22.4$\,\AA\,${\rm MG^{-1}}$ for the strongest blue $\upsigma$ components, but only $+17.1$\,\AA\,${\rm MG^{-1}}$ for the strongest red components, leading in both cases to $\delta B \approx 1.16$\,MG. The estimated spread of $B$ values is then found to be $B \pm \delta B = 11.5 \pm 1.2$\,MG. This value is in agreement with previous estimates of spread in field strength values over the stellar surface. \\

\begin{table*}

\caption{\label{tab:fld_meas} Magnetic field measurements of GD356 using H\,$\upalpha$ and H\,$\upbeta$.\label{tab:Tab4}} 
\begin{tabular}{cccccccccccccc} 

\hline 

JD$-2450000$      
&Phase 
&$\uplambda_{\rm b}$ 
&$\uplambda_{\rm r}$ 
&$\langle | B |\rangle$ 
&$\Delta \uplambda_{\rm Z}$ 
&$\langle B_{\rm z} \rangle$ 
&$W_{\rm \uplambda b}$ 
&$W_{\rm \uplambda r}$\\ 

(d) 
&
&(\AA)     
&(\AA)         
&MG       
&(\AA)    
&(MG)           
&(\AA)       
&(\AA)\\ 

 \hline
 
{\bf 8766.357} 	
&{\bf Avg}  	
&{\bf 6317.14}\,$\pm$\,{\bf 0.86} 		
&{\bf 6776.93}\,$\pm$\,{\bf 0.76} 		
&{\bf 11.50}\,$\pm$\,{\bf 0.03} 	
&{\bf 55.70}\,$\pm$\,{\bf 2.74} 	
&{\bf 2.79}\,$\pm$\,{\bf 0.14} 	
&{\bf 2.86}\,$\pm$\,{\bf 0.07} 	
&{\bf 2.32}\,$\pm$\,{\bf 0.06}\\

8766.317	&0.776   	&$6311.51\pm2.76$ 		&$6773.77\pm2.13$ 		&$11.56\pm0.09$ 	&$58.57\pm7.62$ 	&$2.93\pm0.38$ 	&$2.86\pm0.22$ 	&$2.74\pm0.21$\\ 
8766.324 	&0.863   	&$6321.19\pm2.47$ 		&$6784.60\pm2.42$ 		&$11.59\pm0.09$ 	&$36.73\pm8.66$ 	&$1.84\pm0.43$ 	&$3.11\pm0.21$ 	&$2.18\pm0.19$\\ 
8766.331 	&0.950   	&$6318.21\pm2.90$ 		&$6774.98\pm2.19$ 		&$11.43\pm0.09$ 	&$63.51\pm9.11$ 	&$3.18\pm0.46$ 	&$2.56\pm0.21$ 	&$2.37\pm0.19$\\ 
8766.338 	&0.038   	&$6316.74\pm2.43$ 		&$6777.79\pm2.24$ 		&$11.53\pm0.08$ 	&$65.45\pm8.66$ 	&$3.27\pm0.43$ 	&$3.16\pm0.21$ 	&$2.39\pm0.19$\\ 
8766.346 	&0.137   	&$6317.17\pm2.31$ 		&$6774.31\pm2.18$ 		&$11.43\pm0.08$ 	&$42.14\pm7.88$ 	&$2.11\pm0.39$ 	&$3.63\pm0.23$ 	&$2.69\pm0.21$\\ 
8766.353 	&0.224   	&$6324.37\pm2.97$ 		&$6775.31\pm2.38$ 		&$11.28\pm0.10$ 	&$47.69\pm9.09$ 	&$2.39\pm0.45$ 	&$2.84\pm0.23$ 	&$2.52\pm0.22$\\ 
8766.360 	&0.312   	&$6315.75\pm3.19$ 		&$6776.97\pm2.33$ 		&$11.54\pm0.10$ 	&$51.17\pm8.66$ 	&$2.56\pm0.43$ 	&$2.65\pm0.23$ 	&$2.60\pm0.22$\\ 
8766.367 	&0.399   	&$6322.98\pm3.32$ 		&$6774.14\pm3.05$ 		&$11.28\pm0.11$ 	&$64.79\pm9.98$ 	&$3.24\pm0.50$ 	&$2.69\pm0.25$ 	&$2.08\pm0.23$\\ 
8766.374 	&0.486   	&$6312.64\pm3.11$ 		&$6776.17\pm2.47$ 		&$11.59\pm0.10$ 	&$35.08\pm9.11$ 	&$1.75\pm0.46$ 	&$2.95\pm0.26$ 	&$2.70\pm0.24$\\ 
8766.382 	&0.585   	&$6316.77\pm3.31$ 		&$6781.08\pm3.30$ 		&$11.61\pm0.12$ 	&$55.90 \pm11.27$ 	&$2.80\pm0.56$ 	&$2.97\pm0.27$ 	&$2.12\pm0.25$\\ 
8766.389 	&0.673   	&$6314.37\pm3.82$ 		&$6778.68\pm3.61$ 		&$11.61\pm0.13$ 	&$74.07 \pm11.52$ 	&$3.71\pm0.58$ 	&$2.52\pm0.27$ 	&$2.02\pm0.27$\\ 
8766.397 	&0.772   	&$6312.33\pm3.84$ 		&$6779.07\pm4.95$ 		&$11.67\pm0.16$ 	&$83.90 \pm16.46$ 	&$4.20\pm0.82$ 	&$2.67\pm0.29$ 	&$1.61\pm0.29$\\ 

\hline

{\bf 8766.357}	
&{\bf Avg}	
&{\bf 4688.50}\,$\pm$\,{\bf 0.86} 		
&{\bf 4938.65}\,$\pm$\,{\bf 2.08} 		
&{\bf 11.58}\,$\pm$\,{\bf 0.10}	
&{\bf 36.83}\,$\pm$\,{\bf 1.55} 	
&{\bf 3.41}\,$\pm$\,{\bf 0.14} 	
&{\bf 1.63}\,$\pm$\,{\bf 0.05} 	
&{\bf 2.24}\,$\pm$\,{\bf 0.05}\\

8766.317	&0.776   	&$4685.77\pm4.85$ 		&$4942.55\pm8.07$ 		&$11.89\pm0.44$ 	&$52.87\pm7.09$ 	&$4.90\pm0.66$ 	&$0.94\pm0.17$ 	&$1.95\pm0.16$\\ 
8766.324 	&0.863   	&$4683.62\pm2.86$ 		&$4941.50\pm8.16$ 		&$11.94\pm0.40$ 	&$43.79\pm6.49$ 	&$4.06\pm0.60$ 	&$1.52\pm0.16$ 	&$1.85\pm0.15$\\ 
8766.331 	&0.950   	&$4686.60\pm2.80$ 		&$4942.78\pm8.17$ 		&$11.86\pm0.40$ 	&$35.59\pm5.92$ 	&$3.29\pm0.55$ 	&$1.52\pm0.16$ 	&$1.81\pm0.15$\\ 
8766.338 	&0.038   	&$4690.31\pm3.10$ 		&$4938.30\pm6.74$ 		&$11.48\pm0.34$ 	&$51.90\pm5.33$ 	&$4.81\pm0.49$ 	&$1.40\pm0.16$ 	&$2.26\pm0.15$\\ 
8766.346 	&0.137   	&$4692.29\pm3.22$ 		&$4939.50\pm8.66$ 		&$11.45\pm0.43$ 	&$44.53\pm6.82$ 	&$4.12\pm0.63$ 	&$1.48\pm0.17$ 	&$1.91\pm0.17$\\ 
8766.353 	&0.224   	&$4690.16\pm3.56$ 		&$4936.64\pm7.08$ 		&$11.41\pm0.37$ 	&$35.06\pm5.99$ 	&$3.25\pm0.55$ 	&$1.35\pm0.18$ 	&$2.35\pm0.17$\\ 
8766.360 	&0.312   	&$4686.94\pm2.38$ 		&$4933.98\pm6.30$ 		&$11.44\pm0.31$ 	&$29.73\pm4.38$ 	&$2.75\pm0.41$ 	&$2.02\pm0.18$ 	&$2.65\pm0.17$\\ 
8766.367 	&0.399   	&$4684.60\pm2.49$ 		&$4936.34\pm6.67$ 		&$11.65\pm0.33$ 	&$30.27\pm4.53$ 	&$2.80\pm0.42$ 	&$1.98\pm0.18$ 	&$2.55\pm0.17$\\ 
8766.374 	&0.486   	&$4694.92\pm3.82$ 		&$4937.59\pm6.33$ 		&$11.23\pm0.34$ 	&$36.66\pm4.82$ 	&$3.39\pm0.45$ 	&$1.29\pm0.18$ 	&$2.73\pm0.18$\\ 
8766.382 	&0.585   	&$4691.30\pm2.31$ 		&$4939.30\pm6.11$ 		&$11.48\pm0.30$ 	&$23.12\pm4.24$ 	&$2.14\pm0.39$ 	&$2.16\pm0.18$ 	&$2.86\pm0.18$\\ 
8766.389 	&0.673  	&$4687.08\pm2.28$ 		&$4937.84\pm8.27$ 		&$11.61\pm0.40$ 	&$50.69\pm5.15$ 	&$4.69\pm0.48$ 	&$2.16\pm0.18$ 	&$2.07\pm0.17$\\ 
8766.397 	&0.772   	&$4684.02\pm3.36$ 		&$4936.97\pm7.87$ 		&$11.71\pm0.40$ 	&$30.70\pm5.60$ 	&$2.84\pm0.52$ 	&$1.45\pm0.18$ 	&$2.19\pm0.17$\\ 

\hline 

\end{tabular}
\end{table*}

The data can be further examined for evidence of magnetic field variability during the rotation period of GD\,356. A first assessment of variations may be obtained from Fig.~\ref{fig:Fig1}. This figure suggests strongly that rotational variations do occur. Variability is particularly apparent in the red H\,$\upbeta$ $\upsigma$ lines, in which two of the strongest magnetically split components are approximately stationary with changes of field strength, thus producing narrow emission features in which small changes (for example in amplitude) are particularly obvious to the eye.  In contrast, the blue $\upsigma$ component of H\,$\upbeta$ produces wide observable features because the wavelength vs. field strength dependence is stronger, and variability is not at all obvious by eye. 

In addition the, $\uppi$ components of H\,$\upbeta$ include relatively strong transitions that are nearly but not quite stationary, and so also lead to sharp features in which variations are easier to see than in broad smooth features. In the sharp features of both the $\uppi$ and red $\upsigma$ components of H\,$\upbeta$, clear variability is visible. If one assumes that blue-shifted excursions of the $\uppi$ component at about 4825\,\AA \ are due to increasing field strength (see Fig.~\ref{fig:Fig2}), then it appears from Fig.~\ref{fig:Fig1} that the $\langle | B | \rangle$ may be largest around phase 0.9 and weakest about phase 0.4.

\begin{figure}
\includegraphics[width=\columnwidth]{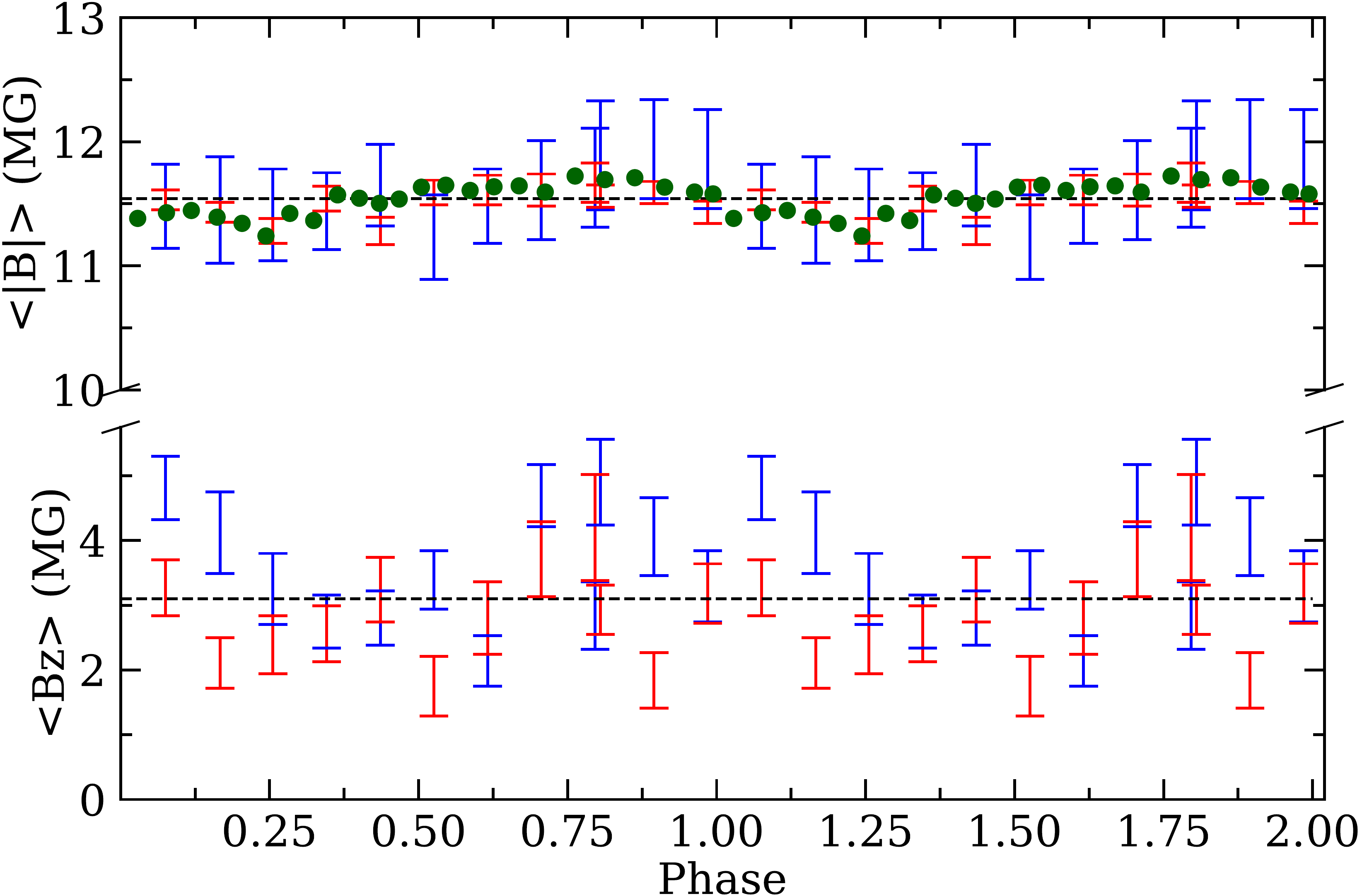}
\vskip 1mm
\caption{$\langle B_{\rm z} \rangle$ and $\langle | B | \rangle$ as functions of \textit{TESS} photometric phase. The spectropolarimetry measurements obtained using H\,$\upalpha$ are shown in red symbols; those from H\,$\upbeta$ are in blue. Dashed black lines represent the average value. In green are shown analogous results from the non-polarimetric spectroscopy, based on the measurements plotted in Fig,~\ref{fig:Fig3}, where a small offset of $+0.28$\,MG has been applied to account for calibration disparities between the two datasets. Each data point is the phase average of approximately 15 exposures in total, and the error bars are roughly the same size as the symbols. The scale of the two $y$ axes are different, and highlight the greater precision of the $\langle | B | \rangle$ measurements.}
\label{fig:Fig8}
\end{figure}

In Fig.~\ref{fig:Fig8}, the \bs\ and \bz\ values are shown as functions of the photometric period and zero point deduced from the \textit{TESS} observations discussed in Sections \ref{sec:obs} and \ref{sec:res}. This figure illustrates two important  points. First, the \bs\ values measured with H\,$\upalpha$ and H\,$\upbeta$ are in excellent agreement; the deduced standard errors seem to be about the right size. Secondly, while the mean field modulus appears to be approximately constant, there does appear to be a low-amplitude variation, possibly 0.1--0.2\,MG in magnitude. This variation is minimum around phase 0.3, and maximum near phase 0.8--0.9, in broad agreement with the indications from Fig.~\ref{fig:Fig1} that the splitting of the H\,$\upbeta$ $\uppi$ component is largest around phase 0.9.

In contrast, the \bz\ measurements obtained from the two Balmer lines are more discrepant than would be expected from the computed uncertainties. This may be due to the pernicious effect of relatively small errors in the flat field corrections applied to the shallow Zeeman components of the emission lines, which are not included in the estimated uncertainties. Apart from this potential problem, the value of $\langle B_{\rm z} \rangle$ is approximately constant around 3.0--3.2\,MG. In fact, the mean longitudinal field is probably weakly variable, again with a minimum value around phase 0.4 and a maximum (or perhaps a double maximum) around phase 0.8 or 0.0. Like the variation suggested by the mean field modulus \bs\ (Fig.~\ref{fig:Fig8}), this variation is suggestive but not completely convincing. 

\section{Discussion}\label{sec:dis}

The following section discusses the findings of all measurements obtained to date, including inferences that can be made about the magnetic field and emission regions of the prototype DAHe star. Ample discussion is dedicated to the only model considered viable prior to this study, the unipolar inductor \citep{Li1998}. This model is re-examined in the context of observations of GD\,356 and what is known from the two recently discovered DAHe stars SDSS\,J125230.93-023417.7 (J1252 hereafter; \citealt{Reding2020}) and SDSS\,J121929.45+471522.8 (J1219 hereafter; \citealt{Gansicke2020}). A few issues that might cause this model to fail from a theoretical perspective are discussed, and an alternative hypothesis for the DAHe subclass is given based on collective stellar properties that make extrinsic mechanisms unlikely.

\subsection{Implications of the Photometric and Spectroscopic Variability}

All data and periodicities identified to date are consistent with a single period previously identified as stellar rotation \citep{Brinkworth2004}. In addition to the broadband flux, the strength of the emission lines and their central wavelength both vary on the same period. The emission feature $W_\uplambda$ values vary in anti-phase with the \textit{TESS} and ground-based photometry, suggesting that either the emission features intrinsically diminish at the photometric maximum, or the emission remains relatively constant and the $W_\uplambda$ variability is produced by continuum brightness fluctuations. However, the latter can be ruled out by pointing out that the continuum flux variation is at most on the order of $\pm1$\, per cent (see Table \ref{tab:Tab3}), while the line fluxes vary by roughly $\pm30$\,per cent.

It is important to emphasise that J1252 appears to exhibit the same anti-phase behaviour where the emission features manifest their maximum strength at the photometric minimum \citep{Reding2020}. There is not yet phase-resolved spectroscopy of J1219, but of the two spectra obtained for this star, it seems to also follow this pattern where emission is strongest when the total optical flux is lowest \citep{Gansicke2020}. To avoid confusion, it is worth noting that for GD\,356 and J1252 phase $=0$ is defined at the light curve maximum, whereas for J1219 the authors use the opposite convention. Thus any successful model should account for the anti-phase behaviour, as it appears intrinsic to the DAHe population to date.

Detected morphological changes in the $\uppi$ components of H\,$\upbeta$ and potentially H\,$\upalpha$ (Fig.~\ref{fig:Fig1}) appear periodic and follow the spin period, suggesting that rotation is responsible. Transformations in the emission profiles are consistent with a modest changes in the magnetic field across the surface as it rotates in and out of the view (more on this in Section 4.2). The broad-band photometric variability is clearly dominated by changes in the photospheric continuum (Table \ref{tab:Tab3}), and is almost certainly due to some type of magnetic field-dependent opacity, as seen in other high-magnetic field white dwarfs, such as Feige\,7 or G183-35 \citep{Achilleos1992,Kilic2019}.

Previous modelling of spectropolarimetric and photometric data independently suggest a 0.1 covering fraction for the 1) polarised emitting region \citep{Ferrario1997a}, and 2) a dark star spot \citep{Brinkworth2004}. These two inferences can be reconciled with the anti-phase behaviour observed in GD\,356 and other DAHe stars if a dark photospheric spot sits below an optically thin emission region, i.e.\ a chromosphere. These two regions -- one hot, one cool -- cannot be spatially independent or the anti-phase behaviour would not be observed. In terms of the unipolar inductor model, there is no a priori reason why a dark surface spot caused by magnetic dichroism should be coincident with an emission region caused by accretion along a current loop from an orbiting companion. There are no such additional signals in any data obtained to date, and despite the independent variations in broad-band flux, emission line strength, and emission line wavelength, only the spin period manifests with confidence.

\subsection{Interpretation of the Magnetic Measurements}

Here, the possible interpretations of the magnetic measurements are considered. If the emission region is localised on the surface of the white dwarf, with a magnetic field that is roughly uniform in strength and approximately vertical in direction over the emission spot, then the small ratio of $\langle B_{\rm z} \rangle / \langle | B | \rangle \approx 0.25 - 0.30$ suggests that the line of sight towards the spot is inclined at a large angle of perhaps $70\degr$ to the vertical direction. In other words, the spot is near the limb of the star as seen from Earth. In this case, for the spot to appear almost unchanging under rotation, the spot must be nearly centred on one of the two poles of stellar rotation, and thus the inclination angle of the rotational axis is similarly large. If correct, this implies the emission region itself is not responsible for the broad-band photometric variations, and consistent with the findings given in Table \ref{tab:Tab3}. 

However, this picture depends strongly on the specific model of the emission as arising in a localised spot on the white dwarf. The localised spot model is supported by observations of the two other DAHe stars, J1252 and J1219, both of which exhibit strong rotational variations in the strength of the magnetised emission region \citep{Reding2020,Gansicke2020}. In fact, the emission region of J1252 appears to vanish over the stellar limb for part of the rotation cycle, indicating that the emission is confined to a limited region of the stellar surface, while an H\,$\upalpha$ absorption line region, in which the line may well be broadened by a weaker magnetic field, extends over a larger part of the white dwarf. 

In a scenario where the distribution of emission over the surface of GD\,356 is different from that deduced from the other two DAHe stars, it is still possible that emission arises from the bulk of the visible stellar disk. The observed, and limited, 1\,MG dispersion of $|B|$ indicates the whole emission region has a nearly uniform value of $|B|$. This has been proposed as an argument against a widely distributed emission region \citep[e.g. by][]{Ferrario1997a} on the grounds that an assumed global dipole magnetic field varies in strength by a factor of two between pole and equator. Nevertheless, little is known about the real global field distributions over magnetic white dwarfs, where only a few models based on phase-resolved magnetic measurements are available in the literature (e.g.\ \citealt{Euchner2005, Valyavin2005, Euchner2006, Valyavin2008}). And in fact for WD\,2047+372, in which the two magnetic hemispheres become independently visible as the star rotates, the value of $\langle | B | \rangle$ remains virtually constant as the star rotates \citep{Landstreet2017}.

Exploring another possibility, if the magnetic emission region on GD\,356 extends over much or all of the stellar surface, then the low-amplitude of variation of $\langle B_{\rm z} \rangle$ suggests that the field structure is close to, but not precisely, axisymmetric about the rotation axis, and that the small value of $\langle B_{\rm z} \rangle / \langle | B | \rangle $ merely reflects the variation in local field vector inclination over the visible stellar surface. One might suspect that the rotational axis makes a relatively small angle (say of the order of $30\degr$) to the line of sight, as it seems probable that field cancellation over the surface in such a global model would reduce the mean projected field to a fairly small fraction of the value of the mean field modulus. 

Despite all this, any interpretation of the nearly constant mean longitudinal field measurements implicitly assumes a production model for the line polarisation in the emission region. This model supposes that the polarisation in the line is produced essentially only in the emission region of inverted temperature and source function gradients, without any significant contribution from the underlying photosphere with a normal, negative temperature gradient. The spectroscopy of J1252 shows that for at least one DAHe star, the flux from the emission line region adds to flux emerging from a surrounding or underlying hydrogen-rich photosphere, producing absorption lines that seem to be broadened by a magnetic field comparable in strength to that in the emission line region  \citep{Reding2020}.

If a region of Balmer absorption lies beneath the emission line region of GD\,356, and shares the local magnetic field, the emergent flux from that absorption region into the overlying emission region will probably already carry a polarisation signature qualitatively like that of the emission lines, but of opposite sign. This should then be further polarised by the overlying emission, but the net emergent polarisation may be rather different than would emerge from the emission line region alone. This might, for example, reduce the net level of polarisation in $\upsigma$ components, and lead to deduced values of $\langle B_{\rm z} \rangle$ that are substantially smaller than the physically correct values. 

\subsection{Challenges to Unipolar Inductor Model}

GD\,356 has defied conventional models and here the reasons are briefly described. Because the emission cannot arise from the cool photosphere with a negative temperature gradient, a heating mechanism is required. Perhaps the most conventional model would be accretion from a mass-transferring companion, but ultraviolet through mid-infrared photometry rules out all unevolved companions down to 12\,M$_{\text{Jup}}$ \citep{Wickramasinghe2010}. Furthermore, both radio observations and a deep X-ray pointing provide strong constraints on accretion from the interstellar medium and photospheric heating by a stellar corona, where the latter observations set an upper limit of $L_X < 6 \times10^{25}$\,erg\,s$^{-1}$, nearly two orders of magnitude lower than the Balmer line emission luminosity \citep{Greenstein1985,Ferrario1997a,Weisskopf2007}. 

Owing to these empirical challenges, the unipolar inductor model has been applied to GD\,356 \citep{Li1998}. In this picture, the surface hot spot is electromagnetically induced by the orbital motion of a conducting planet through the stellar magnetosphere, similar to the Jupiter-Io system \citep{Piddington1968,Goldreich1969}. An exoplanet in a sufficiently close orbit can induce a current loop between itself and the star, resulting in ohmic dissipation and heating in the stellar atmosphere. In principle, the induced potential and dissipated power could be significantly higher than in the Jupiter-Io system. From an observational point of view, the line of sight to the heated region is modulated by stellar rotation, and an additional displacement on the stellar surface due to the motion of the planet through the magnetosphere. The displacement of the spot should be periodic on the orbital period, and possibly further modulated by misalignment of stellar magnetic, rotational, and planetary orbital axes.

Despite the attractions of this model, especially in light of the ubiquitous nature of planetary systems orbiting white dwarfs, there are at least two fundamental flaws when applied to isolated stars such as GD\,356. The first and perhaps less critical problem with the unipolar inductor is that there may be no actual magnetosphere associated with an isolated magnetic white dwarf. That is, there is no a priori reason to expect available ions to form a current sheet and thus enable the inductor to work in the first place. \citet{Li1998} speculate that the ions may be provided either by the interstellar or interplanetary medium, but that is far from certain, especially as the three published DAHe stars are all within the Local Bubble where little interstellar material is to be found. Even if present, it is not clear that passing interstellar material would be ionised and then incorporated into a stellar magnetosphere. While it may be attractive to invoke an interplanetary medium, it is worth pointing out that the heliosphere and some planetary magnetospheres (e.g.\ Mercury, Earth) are powered by plasma from the solar wind. Thus for the unipolar inductor to work at a white dwarf, it requires either a planet be {\em conducting and a source of ions}, or else these two requirements must be met independently.

The second and most damaging problem for the unipolar inductor is stellar rotation, and within this framework there are two distinct ways in which the model can break down. In order for a steady DC circuit to be established (and form a single, localised spot) the Alfv\'en travel time between the planet and star has to be small compared to the time it takes the magnetic field lines (flux tube) to rotate past the companion \citep{Goldreich1969}. This is a point of failure in the Jupiter-Io system due to the rapid planetary rotation, and this results in the appearance of multiple spots and leading and trailing emission regions on either side of the main Io spot (\citealt{Bonfond2013} and references therein). The Alfv\'en speed is expected to be substantially larger for GD\,356 because of its large magnetic moment \citep{Willes2005}, and for an Earth-sized planet orbiting just outside the Roche limit at 1\,R$_{\odot}$, a flux tube requires only 20\,s to rotate past the entire planet diameter in this best case scenario. In the model outlined by \citet{Li1998}, where the planet is in a wider orbit of 10\,h then the flux tube sweeps past the planet in just over 4\,s. Assuming a 13\,MG dipole field (admittedly uncertain, see previous Section) with a magnetospheric plasma of the same number density as the Io torus, an Alfv\'en speed faster than light is obtained, and the Alfv\'en mode practically becomes an electromagnetic mode, travelling at the speed of light. The resulting, round trip Alfv\'en travel time between the white dwarf and the magnetically linked companion will be close to 15\,s just outside the Roche limit, and over 30\,s for the planetary orbit modelled by \citet{Li1998}. Under such conditions, the system marginally fits the DC unipolar conditions if the companion is adequately large and orbits close to the Roche limit, but otherwise fails.

Another critical aspect of rapid rotation is the ability of charge carriers to flow along the flux tube when they experience significant centrifugal forces. The balance between plasma pressure gradient and centrifugal force components parallel to magnetic field lines gives rise to a characteristic plasma length scale given by \citep{Caudal1986}:

\begin{equation}
	\ell = \sqrt{\frac{(Z+1)kT}{m_{\rm i}\upomega^2}}
	\label{Eqn2}
\end{equation} 

\smallskip
\noindent
where $Z$ is the ion charge, $T$ is the plasma temperature, $m_{\rm i}$ is the ion mass, and $\upomega$ is the rotational frequency. The ratio of the plasma scale length $\ell$ to the orbital radius $a$ of the conducting body gives an indication of the efficiency of the resulting current carriers. Taking $T\sim10^5$\,K and singly ionised sulphur for the Jupiter-Io system, $\ell/a\approx0.1$ and thus a hotter plasma source (possibly with lighter ions) is suspected of creating the Io footprint in the aurora of Jupiter, as the unipolar inductor is inefficient. In this context, it is noteworthy that there is a requirement of significant particle acceleration by Alfv\'en waves, based on the observed and modelled relationship between decametric radio emission and the auroral emission of the Io footprint \citep{Zarka1998}. Furthermore, the brightness of the main auroral emission at Jupiter also requires significant particle acceleration in the magnetospheric plasma to provide the required power input (e.g.\ \citealt{Ray2009}), and that recent in situ plasma observations by the {\em Juno} spacecraft have revealed at least two different mechanisms for this particle acceleration \citep{Mauk2018}. Thus, the elucidation of this acceleration process, even at Jupiter, is an ongoing area of investigation.

The situation is made worse when applied to GD\,356 and the DAHe stars because of even more rapid rotation and thus centrifugal forces on ions. In a simplified picture where the plasma is only (hydrogen) protons and electrons, and $T\sim10^5$\,K, and where again a solid planet is just outside the Roche limit at roughly 1\,R$_{\odot}$, the substantially faster rotation speed of GD\,356 yields $\ell/a\approx0.06$ and is not favourable for efficient charge transport. But the model completely breaks down with J1252 which rotates every 317\,s \citep{Reding2020}, and gives a ratio of $\ell/a\approx0.003$. Thus, it appears the unipolar inductor cannot work efficiently in the presence of the likely extreme differential rotation of the star compared to any planetary orbit, as {\em these values are upper limits from the most closely orbiting planet possible}.

Lastly, the unipolar inductor predicts a single surface spot which is heated, rather than the conventional star spot which is cooler than the surrounding photosphere owing to the magnetic suppression of neighbouring convective cells. Thus the anti-phase behaviour of the light curve and emission-line strength in GD\,356 and other DAHe stars does not appear consistent with this simple prediction. The results of the magnetic field measurements indicate the hot region is likely near one of the two {\em rotational} poles, and that is also not necessarily expected if the spot is due to an orbiting planetary body, where instead an unipolar-induced footprint should be near the {\em magnetic} pole, and these two may or may not coincide.

For all the above reasons, the unipolar inductor does not appear to be a successful description of these intriguing white dwarfs from either an observational or theoretical perspective.

\subsection{Intrinsic Chromospheric Activity in White Dwarfs}

It has been independently noted, as mentioned also in \citet{Gansicke2020}, that the three confirmed DAHe stars all lie within a compact region of the Hertzsprung-Russell (HR) diagram. A brief summary of these stars is as follows, and it is notable that all but GD\,356 were reported only in the past year. The second example of this subclass is J1252, and with $B\approx5$\,MG and a 317\,s photometric period, it is currently the fastest-rotating, isolated (magnetic) white dwarf. As previously noted, at some photometric phases the emission lines from J1252 diminish to undetectable levels within complex H\,$\upalpha$ and H\,$\upbeta$ absorption features \citep{Reding2020}, but it is noteworthy that {\em the light curve is consistent with a spot that is never fully out of view}. Most recently published, the optical spectrum of J1219 is superficially similar to GD\,356 but with $B\approx18$\,MG and a 15.3\,h rotation period. All three stars exhibit light curves and emission line strengths consistent with anti-phase variation on the same period.

\begin{figure}
\includegraphics[width=\columnwidth]{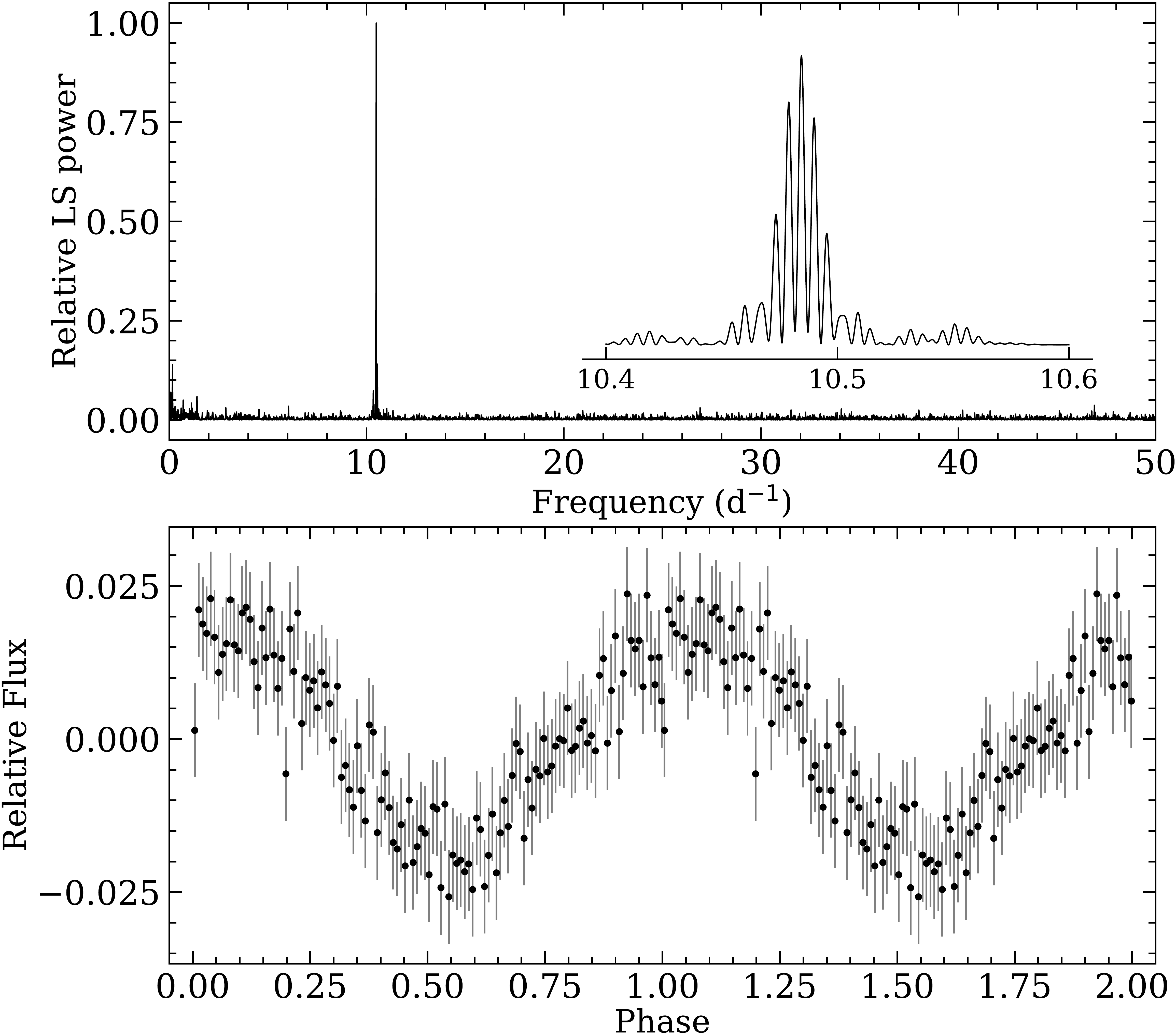}
\vskip 1mm
\caption{\textit{Upper panel}: Normalised LS periodogram of \textit{TESS} photometry of J0412. The peak value corresponds to $10.48437(7)$\,d$^{-1}$. The insert shows a zoomed in version of the periodogram. \textit{Lower panel}: \textit{TESS} photometry folded on the peak period. Each point is an average of 400 recordings. Phase $= 0$ corresponds to the photometric maximum at $T_{\rm BJD}=2458816.031(5)$\,d.}
\label{fig:Fig9}
\end{figure}

Also reported within the past year, there is a fourth DA star with Balmer emission lines, WD\,J041246.85+754942.3 (hereafter J0412; \citealt{Tremblay2020}). It is remarkably similar to the three known DAHe stars; it lies in precisely the same and compact region of the HR diagram, and it has rapid rotation. The {\em TESS} light curve of WD\,J0412+7549 shows photometric modulation with a period of $2.28910\pm0.00002$\,h with amplitude $2.58\pm0.07$\,per cent (Fig.~\ref{fig:Fig9}) and thus potentially signals the presence of spectroscopic variability as seen in GD\,356. This DAe white dwarf exhibits a much weaker emission feature in the core of an otherwise-normal looking H\,$\upalpha$ absorption feature, and has an H\,$\upbeta$ feature whose depth is likely diluted by emission. There are not yet any magnetic field constraints, nor significant changes in three observational epochs \citep{Tremblay2020}, but it is clearly worth further study.

\begin{figure}
\includegraphics[width=\columnwidth]{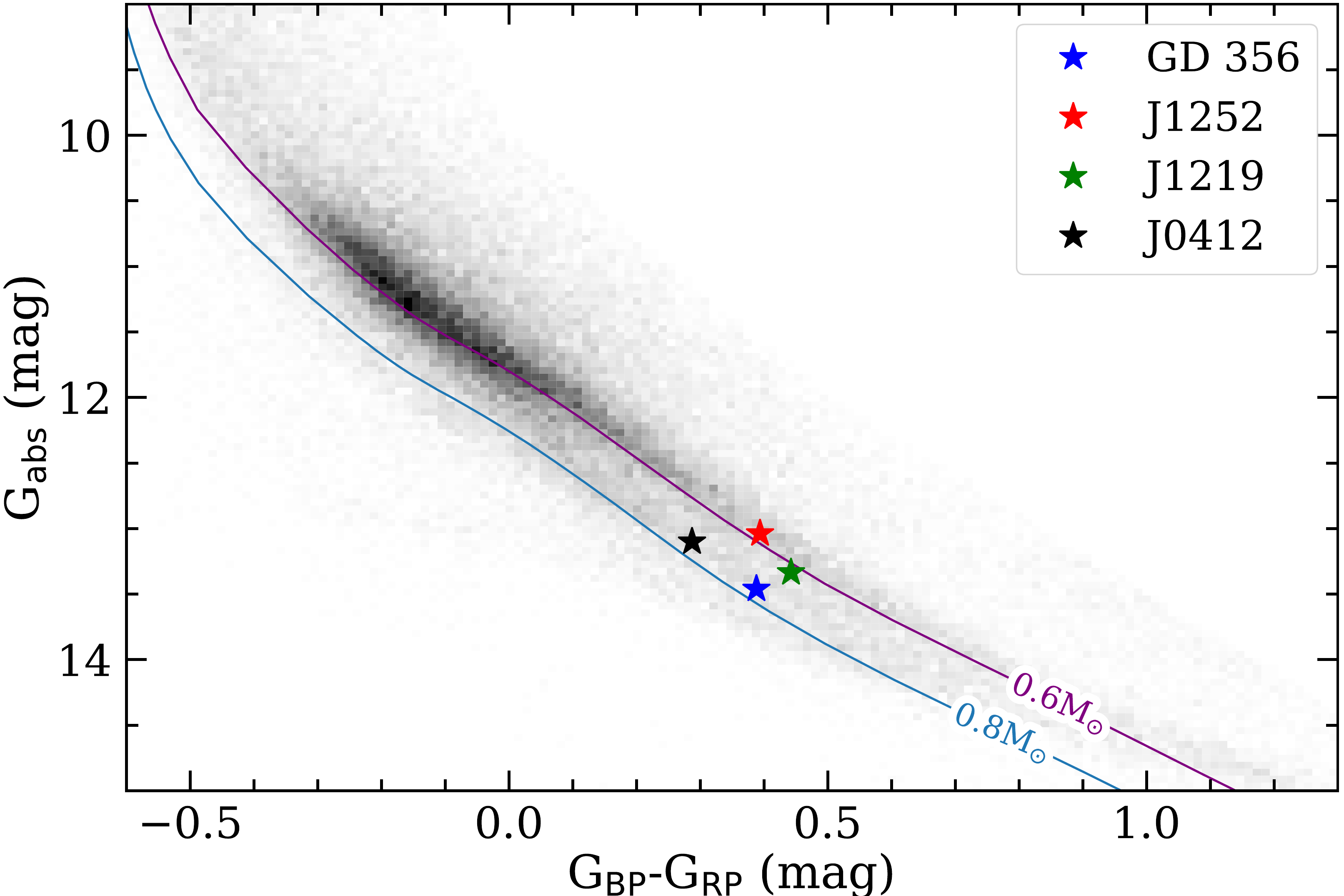}
\vskip 1mm
\caption[pass]
{Magnitude-limited Hertzsprung-Russell diagram of {\em Gaia}-selected white dwarfs with G < 19\,mag \citep{Gentile2019}. Two cooling tracks are shown for pure-hydrogen atmosphere white dwarfs with 0.6 and 0.8 M$_\odot$.\footnotemark By defining a rectangular region with vertices corresponding to the positions of the four known DA(H)e stars, it is estimated that 0.8\,per cent of white dwarfs are enclosed in this region. Therefore, such close clustering of the four objects is highly unlikely to be accidental.}
\label{fig:Fig10}
\end{figure}

From these four stars and their collective and highly similar properties, we hypothesise that the phenomenon of Balmer emission lines in these white dwarfs {\em is an intrinsic stellar property}. In Fig.~\ref{fig:Fig10} all four of these white dwarfs are plotted on an HR diagram, together with a sample of {\em Gaia} DR2-selected white dwarfs with $G< 19$\,mag \citep{Gentile2019}. The position of these DA(H)e stars is truly remarkable, with $<B_{\rm P}-R_{\rm P}> = 0.38\pm 0.07$ and $<M_{\rm G}> = 13.23\pm0.20$\,mag. Given their highly similar $T_{\rm eff}$ estimates, one can take these absolute magnitudes as a decent proxy for their bolometric luminosities, suggesting they are all well within roughly a factor of $\pm2$ in stellar luminosity. This likelihood is substantiated by using the stellar parameter determinations based solely on {\em Gaia} bandpasses and distances \citep{Gentile2019}, where mean stellar luminosities for these four stars appear to be even more tightly correlated with $\langle \log(L_*/L_{\odot}) \rangle = -3.34\pm0.09$\footnote{CL\,Oct, a rapidly rotating DAH white dwarf, has $\log(L_*/L_{\odot})=-2.00$ and is more than $20\times$ intrinsically brighter than the DAHe stars, but may eventually enter an emission phase later in its evolution.}

Their collective properties of these stars include:

\begin{enumerate}

\item{All four stars show emission lines of hydrogen, with no other species in emission.}

\smallskip
\item{All four stars are rotating faster than a typical white dwarf, and in three of four cases the rotation is rapid to extreme.}

\smallskip
\item{All four stars have a similar luminosity, approximated by $M_{\rm G}= 13.2$\,mag and $\log(L_*/L_{\odot})=-3.3$.}

\smallskip
\item{At least three of the stars are highly magnetic with field strengths on the order of MG.}

\end{enumerate}

An analysis of this magnitude-limited white dwarf population can provide a probability of finding all four stars in a narrowly-defined region of the HR diagram. For this analysis, in addition to the $G < 19$\,mag requirement, the following conditions were imposed for white dwarfs to be drawn from the sample \citep{Gentile2019}, in units of magnitude:

\begin{equation}
\begin{aligned}
G_{\rm abs}<10:  G_{\rm abs}-8\times(G_{\rm BP}-G_{\rm RP})>10\phantom{.5}\\
10<G_{\rm abs}<12:  G_{\rm abs}-4\times(G_{\rm BP}-G_{\rm RP})>10\phantom{.5}\\
G_{\rm abs}>12:  G_{\rm abs}-3\times(G_{\rm BP}-G_{\rm RP})> 10.5
\end{aligned}
\end{equation}

\smallskip
\noindent
These cuts effectively remove a portion of white dwarf-main sequence systems, extremely-low-mass white dwarfs and cataclysmic variables. The resulting distribution is shown in Fig.~\ref{fig:Fig10}.

Although such a small number of the DA(H)e stars prohibit meaningful statistical investigation, some inference can be made. Within a quadrangle where each vertex is one of the four emitting stars, there are 0.8\,per cent of all sources. Alternatively, by converting the individual objects into a continuous distribution using kernel density estimation, and sampling it via a Monte Carlo method, it can be shown that such close clustering is rarely achieved. Only about 1.2\,per cent of four randomly sampled objects end up within $\Delta(G_{\rm BP}-G_{\rm RP})<0.31$ and $\Delta G_{\rm abs} < 0.84$ of each other anywhere on the presented HR diagram. These $\Delta$ values correspond to {\em twice the maximum spread} of $G_{\rm BP}-G_{\rm RP}$ and $G_{\rm abs}$ of the four DA(H)e white dwarfs. Even though these methods cannot provide a robust statistical characterisation, it should be nonetheless clear that extrinsic sources are highly unlikely to be the cause of their observed properties.

This fact suggests that these four stars each have an intrinsically activated chromosphere, one that appears dependent on rotation, magnetism, and temperature or luminosity. There are three independent reasons that suggest the observational phenomena are intrinsic. First, as detailed above, if the source of emission were extrinsic, there would be only a tiny chance of finding all four stars in the same position on the HR diagram. Second, if the emission were to become detectable only once these stars become sufficiently faint to unmask the chromosphere, then they would be free to populate the larger portion of the HR diagram. In contrast, the position of the stars in the HR diagram and their highly similar luminosities makes it clear that these stars occupy only a narrow range of phase space.

\footnotetext{\url{http://www.astro.umontreal.ca/~bergeron/CoolingModels/}}
Third, and something that has not yet been noted in the literature, an extrinsic source of matter or ions would be {\em reflected in the composition of the accreted or conducted mass, and also in the chromospheric emission line species} (e.g.\ CVs, convective stars like the Sun). Instead, these stars emit only in Balmer lines with no other species apparent. An interesting counter-example to the DA(H)e stars is the singular case of PG\,1225--079, where the modest Ca\,{\sc ii} H and K emission features seen in a deep and high-resolution spectrum \citep{Klein2011}. The fact that Ca is present in the photosphere together with other heavy elements is well-understood as atmospheric pollution via the accretion of circumstellar, planetary material \citep{Farihi2016}, but in the case of the DA(H)e stars, the lack of other species argues against (planetary) accretion. Therefore, viewed as an intrinsic phenomenon, the Balmer emission lines arise in DAHe stars {\em because their atmospheres are composed of hydrogen}. Thus, the previous inference of a helium atmosphere for GD\,356, based on optical through near-infrared photometry \citep{Bergeron2001}, is suspect. It is well known that highly magnetic white dwarfs have distorted spectral energy distributions, where sometimes not even a single temperature can be confidently assigned \citep{Ferrario2015}.

It is tempting to speculate a bit further based on this potentially new insight into the white dwarf evolution, and include the only other two white dwarfs suspected to have chromospheric emission; the massive DQe stars G227-5 and G35-26. Both stars have emission lines seen only in the ultraviolet, where only the heavy elements C and O are seen towards both stars, with further lines of N (plus Mg and Si possibly) observed from G35-36 \citep{Provencal2005}. If the DAHe stars are emitting from intrinsic chromospheres and all have hydrogen atmospheres, it is tempting to co-identify these two as non-DA counterparts. Historically, DQ stars were modelled with helium atmospheres (e.g.\ \citealt{Bues1973, Grenfell1974, Wegner1984}), but there is now a significant body of evidence that the warmer and hot DQ stars have little or no helium, and are instead bare stellar cores that just avoided detonation as supernovae, and thus have significant carbon and oxygen in their atmospheres \citep{Dufour2007,Dufour2008}. In this picture, the observed emission lines from G227-5 and G35-26 are intrinsic and reflect the composition of the atmosphere, and thus the composition of the outer layers of a stellar core.

Furthermore, the hot DQ stars in particular appear consistent with stellar mergers \citep{Dunlap2015, Cheng2019}, which together with the above paints them as failed type Ia supernovae. But in common with the DAHe stars, many hot DQ stars are rapid rotators \citep{Williams2016}, magnetic \citep{Dufour2010, Dufour2013}, and if so then they share multiple properties. The difference in the stellar parameters of these DQe and DAHe stars, such as luminosity, thus, may be down to their stellar structure and in particular their atmospheric compositions. While speculative, there may be a cluster of DQ(H)e stars awaiting to be found, but may require ultraviolet searches for emission lines. The two known DQe stars have nearly identical $M_G=12.8$\,mag, which is not too dissimilar to the DAHe stars, but total luminosities that are roughly an order of magnitude brighter and not tightly clustered based on {\em Gaia}-derived stellar parameters. Nevertheless, it is an intriguing prospect that intrinsic chromospheres may be present in white dwarfs, and in different regions of the HR diagram based on structure and composition.

In this picture of intrinsic chromospheres for DAHe stars, the following predictions and corollaries result:

\begin{enumerate}

\item{J0412 will exhibit modulation of the emission features on the 2.29\,h rotation period, but in anti-phase.}

\smallskip
\item{J0412 should be magnetic at a detectable level, and spectropolarimetry is ideal to test for this.}

\smallskip
\item{Periodic signals consistent with orbiting planets will not be forthcoming in continued photometric studies of DAHe stars.}

\smallskip
\item{Only species consistent with the stellar atmosphere will be found in emission, regardless of observational sensitivity.}

\smallskip
\item{All four DAHe stars have {\em hydrogen-rich atmospheres}, including GD\,356.}

\smallskip
\item{A search of this region of the HR diagram will find further potential examples.}

\end{enumerate}

In closing, while this hypothesis does not invoke closely orbiting planetary bodies or second generation planets, it does not exclude them. If it is correct that the emission mechanism is intrinsic, then any system hosting planetary material can have potential interactions. For example, if and when a polluted white dwarf passes through this region of the HR diagram, and if it has a favourable magnetic field and rotation, it would be expected to exhibit emission features that reflect the surface compositions. Given that at least 1/3 to 1/2 of isolated white dwarfs host planetary systems \citep{Koester2014}, and thus if the unipolar inductor is not applicable, it does not put any significant dent in the number of evolved planetary systems to characterise in the near future.

\section{Conclusions}\label{sec:conc}

This paper reports periodic variations in the photometry and spectropolarimetry of GD\,356, which are linked to the rotational period. The previously published rotation period of 1.93\,h is in excellent agreement with the latest photometric data. The emission strength and the average magnetic field varies over the rotation, with the emission persisting at all rotational phases. An anti-phase relation between the relative strength of the emission and broad-band photometry has been demonstrated by the new data, and likely present in all known DAHe white dwarfs. By analysing the semi-amplitudes that would be produced from the observed emission variation, it has been shown that the photometric variability must be predominantly due to the continuum flux variation. Also noted are potential morphological dissimilarities in the emission profiles at the same rotational phase in observations separated by ten months, and although the evidence is weak it could be indicative of changes in the emission region. Importantly, \textit{TESS} and LT light curve analysis highlighted no statistically significant photometric modulation or additional signals other than the known spin period, and thus no evidence for a unipolar inductor caused by an orbiting and conducting planetary body.

In the light of the unipolar inductor model, the absence of secondary signals in the photometry does not provide any support to the theory. Moreover, considering the three recently discovered white dwarfs that share a plethora of properties with GD\,356, the likely mechanism behind chromospheric emission appears to be intrinsic. According to an existing model, acoustic waves generated by atmospheric oscillations can potentially trigger chromospheric activity with the predicted luminosity change of less than one percent for a typical white dwarf surface gravity \citep{Musielak2005}.  Although speculative, because GD\,356 is magnetic, perhaps the temperature inversion is the result of thermal pressure overcoming magnetic pressure, but there are currently no such theoretical models.

Despite all the data accumulated to date, there is not yet a complete and consistent picture of such a potential mechanism. However, there are several testable predictions that could be used to evaluate the intrinsic hypothesis for the emission mechanism, and potentially provide additional DAHe candidates. Further study of this emerging population of white dwarfs will likely shed light on physics behind possible chromospheric activity in these magnetic stars.

\section*{Acknowledgements}
The authors thank the reviewer for feedback on the manuscript. Spectropolarimetry and spectroscopy were obtained at the William Herschel Telescope (operated on the island of La Palma by the Isaac Newton Group), under programme ID P8 in Period 19B and service programme SW2018a34 in Periods 18A and 19A, respectively. This paper includes data collected by the {\em TESS} mission, which is funded by the NASA Explorer Program. The Liverpool Telescope is operated by Liverpool John Moores University, and supported by the Instituto de Astrof\'isica de Canarias, with financial support from the UK Science and Technology Facilities Council. This study was based in part on observations conducted using the 1.8m Perkins Telescope Observatory in Arizona, which is owned and operated by Boston University. This research has made use of the NASA ADS, {\sc iraf, Python, matplotlib, astropy} and {\sc lightkurve}. NW was supported by a UK STFC studentship hosted by the UCL Centre for Doctoral Training in Data Intensive Science. JJH and AW acknowledge salary and travel support through the {\em TESS} Guest Investigator Program Grants 80NSSC19K0378 and 80NSSC20K0592, as well as the {\em K2} Guest Observer program 80NSSC19K0162. JDL acknowledges financial support from the Natural Sciences and Engineering Research Council of Canada, funding reference number 6377-2016. TRM acknowledges support from STFC grant number ST/T000406/1, and similarly JF acknowledges STFC grant ST/R000476/1.

\section*{Data Availability}

The data underlying this article will be shared upon reasonable request to the corresponding author. ISIS data are available at the Isaac Newton Group Archive which is maintained as part of the CASU Astronomical Data Centre at the Institute of Astronomy, Cambridge.


\bibliographystyle{mnras}
\bibliography{references}

\bsp    
\label{lastpage}
\end{document}